\documentclass{article}
\usepackage{arxiv}

\usepackage[utf8]{inputenc} 
\usepackage[T1]{fontenc}    
\usepackage[hidelinks]{hyperref}     
\usepackage{url}            
\usepackage{booktabs}       
\usepackage{amsfonts}       
\usepackage{nicefrac}       
\usepackage{microtype}      
\usepackage{lipsum}
\usepackage{bm}
\usepackage{graphicx}
\usepackage{subcaption}
\usepackage{amsmath}
\usepackage{placeins}
\usepackage{siunitx}
\graphicspath{ {./images/} }
\newcolumntype{C}[1]{>{\centering\arraybackslash}p{#1}}
\allowdisplaybreaks[4]

\title{Exact Elastodynamic Homogenization of Laminated Composites Revisited}

\author{
  Chunlin Wu \thanks{Corresponding Author}\\
  Shanghai Institute of Applied Mathematics \\and Mechanics
  ,Shanghai University\\
  Shanghai, 200044 \\
  \texttt{chunlinwu@shu.edu.cn} \\
\And
  Huiming Yin \\
  Department of Civil Engineering and \\Engineering Mechanics,
  Columbia University\\
  New York, NY, 10027 \\
  \texttt{yin@civil.columbia.edu} \\
}

\begin{document}
\maketitle
\begin{abstract}
This paper revisits Willis' elastodynamic homogenization (Mechanics of Materials, 41 (2009) 385-393) of periodic laminate composites and uses the equivalent inclusion-based method (EIM) to evaluate the unique effective constitutive properties with eigen-fields. The microstructure-specific Green's function is essential in Willis's exact homogenization method, which was derived from the corresponding periodic laminates. Our re-examination identifies several algebraic differences in the published formulae for the Floquet number and the related effective properties. In parallel, this paper simulates inhomogeneities with a comparison medium containing two eigen-fields, namely eigenstrain and eigen-momentum, to simulate stiffness and mass density mismatch, respectively. Therefore, the general Green's function can be directly used. The exact homogenization method with the updated specific Green's function is compared to the proposed EIM with eigen-fields homogenization scheme, and excellent agreement is obtained. The present paper provides a base to extend the homogenization framework to multi-dimensional Willis-type homogenization. 
\end{abstract}


\section{Introduction}
Willis-type homogenization provides a rigorous framework for characterizing the macroscopic behaviors of elastodynamic composites \cite{Willis1980}. In the framework, Willis \cite{Willis1997} proposed that the averaged stress and momentum are generally coupled to both averaged strain and velocity. For periodic composite materials, one-dimensional (1D) laminated media serve as an important benchmark because several analytical methods can solve the microscopic local elastic fields. The extracted effective properties can be examined without the additional complexities related to multi-dimensional geometries. Therefore, elastodynamic homogenization of 1D laminates \cite{Willis2009, Willis2012} plays an important role in subsequent studies, which first illustrate the non-local, dispersive features of the effective kernels.

As Srivastava \cite{Srivastava2015} summarized in the comprehensive review, Willis constitutive effective relation is non-local and it satisfies the elastodynamic equation on average, which is substantially different from other conventional steady-state models. In analogy with Willis' effective constitutive relation, several works have extended homogenization to multiphysics, including heat transfer \cite{Shmuel2025, Gal2025a} and piezoelectricity \cite{PernasSalomon2020, Muhafra2022}. 

Although the periodic laminate case studies are analytically tractable \cite{Willis2009}, their treatment requires careful enforcement of boundary and interfacial conditions. While revisiting the original derivations and plots \cite{Willis2009,Willis2012}, several differences were found between the re-derived analytical expressions and numerical curves. These discrepancies arise primarily from two aspects: (i) differences in the coefficients and characteristic equation to obtain the Floquet number for the microstructure-specific Green's function; and (ii) the published curves are not reproduced when the normalized frequency stated in the text is applied. Although these differences do not alter the proposed homogenization framework or its conclusions for the originality and contribution of work to the community, they affect qualitative evaluations of the effective kernels, which are usually used as benchmarks for other homogenization methods. Therefore, it is meaningful for us to document these differences and provide independently cross-validated results. 

Willis \cite{Willis2009} proposed the exact homogenization method for a 1D laminate, combining the source-driven concept and Fourier-series analysis to extract exact effective kernels under Bloch-form settings. Although the exact homogenization method provides an exact analytical framework for a 1D laminate, it requires deriving the microstructure-specific Green's function. Consequently, when the laminate microstructure varies, one must re-derive the corresponding Green's function, which makes extension to different laminate configurations less direct. These limitations motivate alternative formulae that allow the Green's function technique to simulate heterogeneity independently of the specific microstructure, so the latter can be defined for a homogeneous comparison medium. 

Willis \cite{Willis1980} extended the polarization method for elastodynamics, in which heterogeneous composites can be simulated by a comparison medium and two eigen-fields, namely the eigenstress and eigen-momentum, for the material mismatch in stiffness and mass density, respectively. The local fields are independent of the choice of the comparison medium, provided the polarization fields are solved consistently with equivalent stress and momentum conditions. Thanks to the polarization method, Milton and Willis \cite{Milton2007} presented a homogenization framework for random weakly heterogeneous composites, in which the authors suggested an iterative solution procedure. When the matrix is selected as the comparison medium, the polarization method coincides with Eshelby's equivalent inclusion method (EIM) \cite{Eshelby_1957, Fu1983, Mikata1990,srivastava2012overall}.  Although different works \cite{Milton2007, srivastava2012overall} use different polarization fields, they satisfy consistent equivalent conditions and therefore recover the same actual local fields. 

The exact homogenization \cite{Willis2009} relies on the specific Green's function satisfying the boundary conditions and periodic geometric properties of the composite domain. Such boundary-specific Green's functions are only available for relatively simple and regular domains, i.e., a unit cell under Bloch-form boundary conditions \cite{srivastava2012overall}. Our recent work \cite{Wu2025} incorporated the EIM into an inclusion-based boundary element method (iBEM) \cite{yin2022inclusion}, in which boundary effects are handled through boundary integral equations. Consequently, we can apply the infinite-domain Green's function in the comparison medium, which avoids the derivation of a boundary-specific Green's function for a certain periodic microstructure. 

Despite advances in solving elastodynamic wave propagation, the effective properties cannot be uniquely defined unless a sufficient set of independent sources is provided. Fietz and Shvets \cite{Fietz2009} first found the nonuniqueness issue. Subsequently, Willis \cite{Willis2011} extended the discussion for electromagnetic wave homogenization and proposed the residual-field method to identify unique effective properties. Based on \cite{Willis2011}, the nonuniqueness arises from two aspects: (i) the de-phased averaged momentum and stress are linearly related through the governing equation; and (ii) under Bloch-form boundary conditions, the de-phased averaged displacement and strain are kinematically constrained, as they are obtained from the same potential. Specifically, Willis derived the unique constitutive relation using residual strain and residual momentum, and the method was discussed in several subsequent works \cite{alu2011prb,nassar2015willis,Sieck2017prb}. Introducing residual fields allows the formation of modified constitutive relations among averaged stress, momentum, displacement, and mechanical strain. Because the averaged displacement and mechanical strain are not kinematically constrained, the cross-coupling terms are captured rather than absorbed into the direct effective kernels. The residual-field method is not only used in elastodynamics, but also extended to heat transfer \cite{Gal2025a} and piezoelectric \cite{PernasSalomon2020, Lee2023}. 

Motivated by the above considerations, this work revisits the exact homogenization of 1D laminates and re-examines several analytical expressions and numerical results reported in the text. Furthermore, we verify the results using the EIM with eigen-field homogenization scheme. Section 2 presents the specific numerical case study, defines average operations, and reviews Willis constitutive relations. Section 3 revisits the numerical example in \cite{Willis2009, Willis2012} and updates the Green's function and figures. Section 4 presents the equivalent-inclusion method formulae for 1D laminates and the eigen-field homogenization scheme. Section 5 compares the two methods and demonstrates the excellent agreement with each other. Finally, we provide some concluding remarks. 

\section{Problem statement}

\subsection{Microstructure and boundary conditions}

\begin{figure}
    \centering
    \includegraphics[width = 0.7 \textwidth]{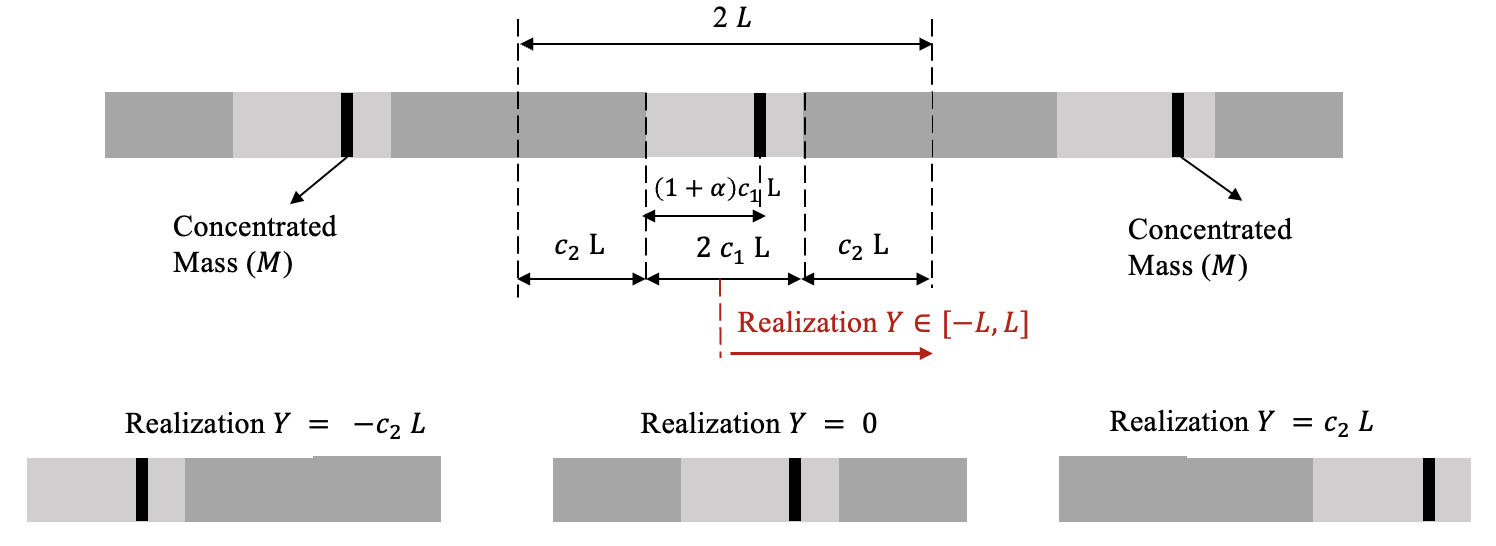}
    \caption{Schematic plot of a periodically laminated composite according to \cite{Willis2009, Willis2012}, and the length of the unit cell is $2 L$. The laminate consists of two materials; the dark gray phase possesses $2 c_{2} L$, while the light gray one occupies $2 c_{1} L$ ($c_1 + c_2 = 1$). A concentrated mass $M = 2 \rho_{1} L$ is introduced to break the reflection symmetry of the unit cell. $Y \in [-L, L]$ refers to the realization variable, which determines the starting and ending of the unit cell. When $Y = 0$, the concentrated mass is placed at $\alpha c_{1} L$, $\alpha = \frac{3}{4}$ in this example.}  
    \label{fig:1d_example}
\end{figure}

Following Willis' original works \cite{Willis2009, Willis2012}, Fig. \ref{fig:1d_example} schematically plots a periodically laminated composite, and the unit cell has the length $2 L$. The laminate consists of two material phases: (i) phase 1 (light gray) possesses $2 c_1 L$; and (ii) phase 2 (dark gray) occupies $2 c_2 L$, where $c_1 = 0.4$ and $c_2 = 0.6$. The two material phases exhibit different elastodynamic properties, in which the stiffness and mass density of the $i^\text{th} (i=1,2)$ phase are denoted as $E_i, \rho_i$, respectively. The stiffness and mass density are different for (i) nonunique case: $E_1 = 0.05, E_2 = 1$, and $\rho_1 = \rho_2 = 0.5$ (defined in Section 4 of \cite{Willis2009}); and (ii) unique case with residual fields \cite{Willis2012}, $E_1 = 0.05 - 0.01i, E_{2} = 1 - 0.1i$ and $\rho_1 = \rho_2 = 0.5$ (defined in Section 5 of \cite{Willis2012}). A concentrated mass $M = 2 \rho_1 L$ is placed at the point $x_p = Y+\alpha c_1 L$ to break the symmetry of the unit cell.

For a periodically laminated composite, the realization variable $Y \in [-L, L]$ determines the starting and ending points of the unit cell, which may alter the effective properties accordingly. With the realization variable $Y$, the local constitutive parameters can be expressed through shifting, $E_Y(x) = E_{Y=0}(x - Y)$ and $\rho_Y(x) = \rho_{Y=0}(x-Y)$. When the realization variable $Y = 0$, the local constitutive parameters can be written as,  
\begin{equation}
    E_{Y= 0}(x) = \begin{cases} E_2 & x\in[-L, -c_1 L) \\ E_1 & x \in [-c_1 L, c_1 L] \\ E_2 & x \in (c_1 L, L] \end{cases} \quad \text{and} \quad \rho_{Y=0}(x) = \begin{cases} \rho_2 & x\in[-L, -c_1 L) \\ \rho_1 & x \in [-c_1 L, c_1 L] \\ \rho_2 & x \in (c_1 L, L] \end{cases} \quad \text{and} \quad \beta_{Y=0} = \alpha c_1 L
    \label{eq:properties} 
\end{equation}
where $\beta_{Y}$ refers to the position of the concentrated mass. When the composite system is under harmonic excitation $\omega$, the spatial and temporal parts can be separated, such as $\overline{\sigma}(x, t) = \sigma(x) \exp[-i \omega t]$, where $\sigma(x)$ refers to the spatially varying part of stress. Similarly, $p(x), u(x)$, and $f(x)$ can be defined for spatial parts of momentum, displacement, and the body force, respectively. Provided the realization variable $Y$, the elastodynamic governing equation in the frequency domain can be constructed as, 
\begin{equation}
    \sigma_{Y,x}(x) + i \omega p_Y(x) + f(x) = 0 
    \label{eq:gov_eqn}
\end{equation}
where $\sigma_{Y}(x)$ and $p_Y(x)$ are stress and momentum with the $Y$ realization; the subscript $(.)_{,x}$ refers to partial differentiation with respect to $x$; and the time partial differentiation of the momentum is represented by the factor $-i \omega$. Note that source force $f(x)$ is prescribed, which is independent of the realization variable $Y$. Accordingly, the local constitutive relations can be written for each realization as, 
\begin{equation}
    \sigma_{Y}(x) = E_{Y}(x) u_{Y,x}(x) \quad \text{and} \quad p_Y(x) = -i \omega \rho_Y(x) u_Y(x)
    \label{eq:local_cons}
\end{equation}
where $u_Y(x)$ and $u_{Y,x}(x)$ are displacement and its gradient for realization $Y$. The unit cell is subjected to Bloch-form boundary conditions. Specifically, the displacement and traction at the starting point are related to those at the ending point through the factor $\exp[2 i \zeta L]$, 

\begin{equation}
    u(L) = u(-L) \exp[2 i \zeta L] \quad \text{and} \quad T(L) = -T(-L) \exp[2 i \zeta L]
    \label{eq:bloch}
\end{equation}
where $u$ and $T$ are displacement and traction, respectively, satisfying $T(x) = E(x) n(x) u_{,x}$, in which $n(-L) = -1$ and $n(L) = 1$ denote the outward direction; $\zeta$ is the macroscopic wavenumber. When $\zeta = 0$, the Bloch-form boundary condition reduces to the periodic boundary condition, and effective properties extracted at $\zeta = 0$ are interpreted as the long-wavelength limit. Although local elastodynamic performance is significant for stress analysis and vibration control, effective properties characterize the composite's overall performance for material design.

\subsection{Definition of averages}

To characterize the effective properties, it is necessary to define the averages of elastodynamic fields. Note that there exist two groups of averages: (i) the spatial average, related to spatial distribution functions; and (ii) the ensemble average associated with probability density functions. Because the unit cell has the length $2 L$, the de-phased spatial average is defined as, 
\begin{equation}
    \langle u \rangle _Y = \frac{1}{2L} \int_{-L}^{L} u_Y(x) \exp[-i \zeta x] w_Y(x) \thinspace dx
    \label{eq:dephase_avg}
\end{equation}
where $\langle . \rangle_Y$ refers to the de-phased average with realization $Y$; $\exp[-i \zeta x]$ is the de-phase factor, which helps to extract the average of the periodic part in the Bloch-form field; $w_Y(x)$ is the spatial weight function with realization $Y$. Note that the de-phased average was also proposed by Srivastava and Nemat-Nasser \cite{srivastava2012overall}. Our recent work \cite{Wu2026_rspa} has demonstrated that the de-phased average is essential to recover Willis reciprocity relations, while effective properties using the conventional average based on the direct domain integral do not satisfy physical constraints. Willis \cite{Willis2009} proposed two typical examples for $w_Y(x)$ that (i) $w_Y(x) = 1$ is a uniform weight function; and (ii) $w_Y(x) = 0$ for the light gray phase (phase 1), and $w_Y(x) = \frac{1}{c_{2}}$ for the dark gray phase (phase 2). 

In addition to the de-phased average, Willis \cite{Willis1997} introduced the ensemble average to evaluate the expectation of a random field over all admissible realizations of the microstructure. Let $\mathcal{P}(Y)$ denote the probability density function of the realization variable $Y$; the ensemble average of the displacement can be defined as, 
\begin{equation}
    \langle u \rangle(x) = \int_{-L}^{L} u_Y(x) w_Y(x) \mathcal{P}(Y) \thinspace dY
    \label{eq:ens_avg}
\end{equation}
where $\langle . \rangle$ refers to the ensemble average and $\mathcal{P}(Y) = \frac{1}{2L}$ is adopted in the present work. The ensemble average and the spatial average in Eq. (\ref{eq:dephase_avg}) therefore operate on different variables. The former averages a field at a fixed position over all realizations, while the latter averages the field within a unit cell with a prescribed realization. In the formulae of \cite{Willis2009, Willis2012}, the averaged stress and momentum are ensemble averages, while the averaged displacement and velocity are weighted ensemble averages. 

\subsection{Willis effective constitutive relation}

For a periodically translated medium, the ensemble average reduces to an integral over all possible translations within one unit cell. Under Bloch-form excitation, the field quantities satisfy $u_Y(x) = \exp[i \zeta Y] u_{Y=0}(x - Y)$ and the weight function satisfies $w_Y(x) = w_{Y=0}(x-Y)$. Therefore, the weighted ensemble average can be expressed as: 
\begin{equation}
    \langle u \rangle(x) = \exp[i \zeta x] \frac{1}{2L} \int_{-L}^{L} u_{Y=0}(s) w_{Y=0}(s) \exp[-i \zeta s] \thinspace ds = \exp[i \zeta x] \langle u \rangle_{Y=0}
\end{equation}
where $\mathcal{P}(Y) = \frac{1}{2L}$, $s = x - Y$.
Hence, the de-phased spatial average $\langle u \rangle_{Y=0}$ is the macroscopic amplitude of the ensemble-averaged fields. Therefore, the effective constitutive properties can be extracted from the de-phased averages for a single realization, without evaluating the ensemble averages over the unit cell. The Willis effective constitutive relations can be written as, 
\begin{equation}
    \begin{bmatrix} \langle \sigma \rangle \\ \langle p \rangle \end{bmatrix} = \begin{bmatrix} E^\text{eff} & S^\text{eff} \\ S^{\dagger, \text{eff}} & \rho^\text{eff} \end{bmatrix} * \begin{bmatrix} \langle u \rangle_{,x} \\ -i \omega \langle u \rangle \end{bmatrix}
\end{equation}
where the operator ``$*$'' refers to the spatial convolution. For a periodically laminated structure subjected to Bloch-form boundary conditions, the ensemble-averaged quantities have the common spatial factor $\exp[i \zeta x]$. Hence, in the Fourier space, taking strain for instance, we can write the ensemble-averaged field quantities explicitly as: 
\begin{equation}
\begin{aligned}
    \widetilde{\langle u \rangle_{,x}} = \frac{ i\zeta}{2\pi} \int_{-\infty}^{\infty} \exp[-i (k - \zeta) x] \langle u \rangle_{Y=0} dx = i \zeta \langle u \rangle_{Y=0} \delta(k - \zeta)
\end{aligned}
\end{equation}
where $k$ refers to the Fourier wavenumber and the other terms including $\langle \tilde{\sigma} \rangle$, $\langle \tilde{p} \rangle$, and $\langle \tilde{u} \rangle$ can be written in the same way. Hence, the Willis constitutive relation in the wavenumber-frequency space can be expressed as, 
\begin{equation} 
\begin{bmatrix} \langle \tilde{\sigma} \rangle \\ \langle \tilde{p} \rangle \end{bmatrix} = \begin{bmatrix} \tilde{E}^\text{eff} & \tilde{S}^\text{eff} \\ \tilde{S}^{\dagger, \text{eff}} & \tilde{\rho}^\text{eff} \end{bmatrix} \thinspace \thinspace \begin{bmatrix} i \zeta \langle u \rangle_{Y=0} \\ -i \omega \langle u \rangle_{Y=0} \end{bmatrix}, \qquad \text{when } k = \zeta
\label{eq:effect_cons} 
\end{equation}
where $(\tilde{.}) (\zeta, \omega)$ refers to effective kernels in the wavenumber-frequency space. Since $ i \zeta \langle u \rangle_{Y=0}$ and $-i \omega \langle u \rangle_{Y=0}$ are kinematically related, extracting effective properties directly from macroscopic responses requires at least two independent sources, such as body force and residual strain \cite{Willis2011}. Specifically, the present work employs the eigenstrain as a residual field to evaluate the unique effective constitutive kernels.

\section{Re-examination of the Green's function in exact homogenization method}

\subsection{Derivation of microstructure-specific Green's function}
The exact homogenization method requires deriving the microstructure-specific Green's function, and Willis \cite{Willis2009} provided an analytical solution of the laminated structure in Fig. \ref{fig:1d_example}. However, re-examining the 1D analytical solution shows the differences in the published Green's function in \cite{Willis2009, Willis2012}, which will be elaborated in the following, and the numerical cross-validations are provided in Section \ref{sec:example}. 

The governing equation with the realization $Y = 0$ with a unit point source can be written as, 
\begin{equation}
    \left[ E_{Y=0}(x) G_{Y=0,x}(x, x') \right]_{,x} + \omega^2 \rho_{Y=0}(x) G_{Y=0}(x, x') = -\delta(x - x')
    \label{eq:green_gov}
\end{equation}

Using the standard theory \cite{Eastham1973}, the solution can be expressed in the Floquet form \cite{Willis2009}, 
\begin{equation}
    G_{Y=0}(x, x') = D \begin{cases} \phi_{+}(x) \phi_{-}(x') & x < x' \\ \phi_+(x') \phi_-(x) & x > x' \end{cases}, \qquad D = \frac{1}{E_{Y=0}(x)} \left[ \phi'_+(x) \phi_-(x) - \phi_+(x) \phi'_-(x) \right]^{-1}
    \label{eq:G_0}
\end{equation}
where $\phi_+(x) = \exp[\mu x] \psi_+(x)$ and $\phi_-(x) = \exp[-\mu x] \psi_-(x)$, and $\psi_\pm(x)$ are periodic over the period $2 L$; $\mu$ is the Floquet number and it characterizes the eigen-mode of the microstructure. Note that although Eq. (\ref{eq:G_0}) shows that the coefficient $D$ is a function of $x$, the resulting coefficient is independent of $x$. Without loss of generality, after we derive the positive branch $\phi_+$ associated with $\mu$, the negative branch $\phi_-$ can be obtained by setting the Floquet number as $-\mu$. 

Within one unit cell, $\phi_+$ can be constructed in terms of four branches, 
\begin{equation}
    \phi_+(x) = \begin{cases}
    \exp[-\mu L] \left( A \cosh[k_2 (L + x)] + B \sinh [k_2 (L + x)] \right) & x \in [-L, -c_1 L) \\ 
    \cosh \left[ k_1 x \right] + b \sinh \left[ k_1 x \right]  & x \in [-c_1 L, \alpha c_1 L) \\
     c \cosh \left[ k_1 x \right] + d \sinh \left[ k_1 x \right] & x \in [\alpha c_1 L, c_1 L)\\
    \exp[\mu L] \left( A \cosh \left[ k_2 (L - x) \right] - B \sinh \left[ k_2 (L - x) \right] \right) & x \in [c_1 L, L]
    \end{cases}
    \label{eq:form_sol}
\end{equation}
where $A, B, b, c, d$ are coefficients to be determined using continuity and jump conditions. To facilitate the derivation of coefficients, we follow Appendix A of \cite{Willis2009} to define some symbols, 
\begin{equation*}
    \begin{aligned}
    & k_1 = -i \omega \sqrt{\frac{\rho_1}{E_1}}, \quad k_2 = -i \omega \sqrt{\frac{\rho_2}{E_2}}, \quad Z_1 = \sqrt{E_1 \rho_1}, \quad Z_2 = \sqrt{E_2 \rho_2} \\ 
    & C_\alpha = \cosh [k_1 \alpha c_1 L], \quad  S_{\alpha} = \sinh [k_1 \alpha c_1 L], \quad C_{1} = \cosh [k_1 c_1 L], \quad S_1 = \sinh [k_1 c_1 L] \\ 
    & C_{1 - \alpha} = \cosh [k_1 (1 - \alpha) c_1 L], \quad S_{1-\alpha} = \sinh [k_1 (1 - \alpha) c_1 L], \quad C_2 = \cosh [k_2 c_2 L], \quad S_2 = \sinh[k_2 c_2 L]
    \end{aligned}
\end{equation*}
Note that the typographical differences in the definition of $C_2$ and $S_2$ are noted here; and $C_1$ and $S_1$ are introduced for notational convenience as the special case of $C_{1-\alpha}$ and $S_{1-\alpha}$ at $\alpha =0$.  

\subsection*{(i) Continuity and jump conditions at the concentrated mass $x = \alpha c_1 L$}

When $x = \alpha c_1 L$, the continuity of two branches in Eq. (\ref{eq:form_sol}) provides, 
\begin{equation}
    C_{\alpha} + b S_{\alpha} = c C_{\alpha} + d S_{\alpha}
    \label{eq:con_alpha}
\end{equation}
The concentrated mass introduces a singular inertia force at $x = \alpha c_1 L$, which causes a jump in the displacement gradient as well as the stress cross the point. The governing equation in the absence of the unit point source becomes, 
\begin{equation}
    E_1 \frac{d^2 \phi_+(x)}{dx^2} + \omega^2 \rho_1 \phi_+(x) + \omega^2 M \phi_+(x) \delta(x - \alpha c_1 L) = 0
    \label{eq:jump_alpha}
\end{equation}
where $M$ is the concentrated mass. Integrating Eq. (\ref{eq:jump_alpha}) over an infinitesimal interval around $x = \alpha c_1 L$, and using the displacement continuity in Eq. (\ref{eq:con_alpha}), we obtain: 
\begin{equation}
    -k_1 E_1 (S_\alpha + b C_\alpha) = -k_1 E_1 (c S_{\alpha} + d C_{\alpha}) - \omega^2 M (C_\alpha + b S_\alpha) 
\end{equation}
which provides coefficients $c$ and $d$, 
\begin{equation}
\begin{aligned}
    c = 1 + \frac{\omega^2 M}{k_1 E_1} S_{\alpha} (C_\alpha + b S_{\alpha}) = 1 + \frac{i \omega M}{Z_1} S_{\alpha} (C_\alpha + b S_\alpha) \\
    d = b - \frac{\omega^2 M}{k_1 E_1} C_{\alpha} (C_\alpha + b S_{\alpha}) = b - \frac{i \omega M}{Z_1}C_{\alpha} (C_\alpha + b S_{\alpha})
\end{aligned}
    \label{eq:cd}
\end{equation}

\subsection*{ (ii) Continuity conditions at two interfaces $x = \pm c_1 L$}
Continuity of displacement and traction across two interfaces provides four additional equations for coefficients $A$ and $B$. The continuity of displacement provides, 
\begin{equation}
    \begin{aligned}
    2A C_{2} = \exp[\mu L] (C_1 - b S_1) + \exp[-\mu L] \left( C_1 + b S_1 - \frac{i \omega M}{Z_1} S_{1 - \alpha} (C_\alpha + b S_\alpha)  \right) \\ 
    2B S_{2} = \exp[\mu L] (C_1 - b S_1) - \exp[-\mu L] \left( C_1 + b S_1 - \frac{i \omega M}{Z_1} S_{1- \alpha} (C_\alpha + b S_{\alpha}) \right)
    \end{aligned}
    \label{eq:con_disp}
\end{equation}
where the above relation obtained through continuity of displacement is identical to Eq. (A.5) of \cite{Willis2009}, and the re-derivation uses $-i \omega$ instead of the variable $s$. 

Subsequently, the continuity of traction provides, 
\begin{equation}
    \begin{aligned}
    2 A S_{2} = \frac{Z_1}{Z_2} \left\{ \exp[\mu L] (b C_1 - S_1) - \exp[-\mu L] \left( b C_1 + S_1 - \frac{i \omega M}{Z_1} C_{1-\alpha} (C_\alpha + b S_{\alpha}) \right)  \right\} \\
    2 B C_2 = \frac{Z_1}{Z_2} \left\{ \exp[\mu L] (b C_1 - S_1) + \exp[-\mu L] \left( b C_1 +  S_1 - \frac{i \omega M}{Z_1} C_{1 - \alpha} (C_\alpha + b S_\alpha) \right) \right\}
    \end{aligned}
    \label{eq:con_trac}
\end{equation}
where the relation obtained from the continuity condition is identical to Eq. (A.6) of \cite{Willis2009}. Because coefficients $A$ and $B$ are uniquely defined for each Green's function, one can obtain two equations for the coefficient $b$ through Eqs. (\ref{eq:con_disp}) and (\ref{eq:con_trac}), 
\begin{equation}
    \begin{aligned}
    b = \frac{(\exp[\mu L] - \exp [-\mu L]) \left( C_1 S_2 + \frac{Z_1}{Z_2} C_2 S_1 \right) - i \omega \exp \left[ -\mu L \right] M C_\alpha \left( \frac{S_{1-\alpha} S_2}{Z_1} +  \frac{C_{1-\alpha} C_2}{Z_2} \right)}{(\exp[\mu L] + \exp [-\mu L]) \left( S_1 S_2 +  \frac{Z_1}{Z_2} C_1 C_2 \right) + i \omega \exp \left[ -\mu L \right] M S_\alpha \left( \frac{S_{1-\alpha} S_2}{Z_1} + \frac{C_{1-\alpha} C_2}{Z_2} \right)} \\
    b = \frac{(\exp[\mu L] - \exp [-\mu L]) \left( C_1 C_2 + \frac{Z_1}{Z_2} S_1 S_2 \right) + i \omega \exp \left[ -\mu L \right] M C_\alpha \left( \frac{S_{1-\alpha} C_2}{Z_1} +  \frac{C_{1-\alpha} S_2}{Z_2} \right)}{(\exp[\mu L] + \exp [-\mu L]) \left( S_1 C_2 + C_1 S_2 \frac{Z_1}{Z_2} \right) - i \omega \exp \left[ -\mu L \right] M S_\alpha \left( \frac{S_{1-\alpha} C_2}{Z_1} + \frac{C_{1-\alpha} S_2}{Z_2} \right)}
    \end{aligned}
    \label{eq:b_coeff}
\end{equation}

\subsection*{(iii) Determination of the Floquet number}
As Eq. (\ref{eq:b_coeff}) indicates, the coefficient $b$ can be expressed in two equivalent forms. Therefore, the Floquet number $\mu$ is determined by requiring the two expressions in Eq. (\ref{eq:b_coeff}) to yield the same value, which forms, 
\begin{equation}
    Q \left(\exp[2 \mu L] + \exp[-2 \mu L]\right) + C = 0, \quad \mu_{\pm} = \frac{\ln \left[ \frac{-C \pm \sqrt{C^2 - 4 Q^2}}{2 Q} \right]}{2L}
    \label{eq:mu}
\end{equation}
where 
\begin{equation*}
\begin{aligned}
    & Q = -Z_1 Z_2\\
    & C = 4 C_1 C_2 S_1 S_2 (Z_1^2 + Z_2^2) + 2 Z_1 Z_2 (C_1^2 + S_1^2) (C_2^2 + S_2^2) - i \omega M \\ & \qquad \times \Big[ \frac{C_2 S_2}{Z_1} \left( (Z_1^2 + Z_2^2) (C_1^2 + S_1^2) + (Z_1^2 - Z_2^2) (C_\alpha^2 + S_{\alpha}^2) \right) + 2 Z_2 (C_2^2 + S_2^2) C_1 S_1 \Big]
\end{aligned}
\end{equation*}
After re-derivation, Eq. (A.9) in \cite{Willis2009} is found to be identical to the present result except for one contribution associated with the concentrated mass. Specifically, algebraic equivalence with the present characteristic equation is obtained when an additional factor of $2$ is included in this term of Eq. (A.9) of \cite{Willis2009}, i.e., 
\begin{equation*}
Z_1 M s \left\{ \cdots \right\} \quad (\text{original})  \rightarrow
2 Z_1 M s \left\{ \cdots \right\} \quad (\text{revised})
\end{equation*}
Since the Floquet number determines the Green's function, including its propagation and all associated coefficients, this difference affects the subsequent evaluations of Fourier coefficient and the resulting effective kernels. 

\subsection*{(iv) Re-evaluation of Fourier coefficients}
Once the Floquet number $\mu$ and the coefficients are determined, the other coefficients $c, d, A$ and $B$ in the piecewise solution can be obtained through the continuity conditions in step (ii). Therefore, the two Floquet solutions with $\mu$ and $-\mu$ can be completely acquired. Their periodic parts, including the stress- and momentum-related fields, can be expressed in terms of Fourier series. Following the definitions of the Fourier coefficients introduced in Eqs. (3.8), (3.21), and (3.25) of \cite{Willis2009}, the required coefficients are obtained by substituting the piecewise Floquet solutions into the corresponding integrals and evaluating them over three intervals, which provide the explicit expressions of $a_n, b_{n}$, and $c_{n}$, which are subsequently utilized in the ensemble-averaged Green's function and effective constitutive kernels. 

Note that the present re-derivation provides the following expression corresponding to the stress-related Fourier coefficients in Eq. (A.13) of \cite{Willis2009}, 
\begin{equation}
\begin{split}
    b_{n}^{(\alpha)} & = k_1 E_1  \Big[ \frac{1 + b}{4} \frac{\exp[\alpha c_1 (i n \pi - \mu L + k_1 L)] - \exp[-c_1 (i n \pi - \mu_L + k_1 L)]}{i n \pi - \mu L + k_1 L} \\ & 
    \qquad \qquad - \frac{1-b}{4} \frac{\exp[\alpha c_1 (i n \pi - \mu L - k_1 L)] - \exp[-c_1 (i n \pi - \mu_L - k_1 L)]}{i n \pi - \mu L - k_1 L} \Big]
\end{split}
\end{equation}
where the re-derivation yields $(1+b)/4$ and $(1-b)/4$, rather than the factors $(c+d)/4$ and $(c-d)/4$ showing in \cite{Willis2009}. And the following equation corresponds to the mass-density-related Fourier coefficients in Eq. (A.14) of \cite{Willis2009}, 

\begin{equation}
    c_{n}(\mu) = \rho_1 [ a_{n}^{(\alpha)}(\mu) + a^{(1-\alpha)}_{n}(\mu) ] + \rho_2 a_{n}^{(2)}(\mu) + \frac{M}{2L} (C_\alpha + b S_\alpha) \exp[\alpha c_1 (i n \pi - \mu L)]
\end{equation}
where the re-derivation yields $(C_\alpha + b S_\alpha)$, rather than the factor $(1+b)$ showing in \cite{Willis2009}. Here, coefficients $c_n$ denotes the Fourier coefficients associated with the mass-density-weighted field, which follows the notation of \cite{Willis2009}. It should not be confused with the coefficient $c_{1}, c_{2}$ in the volume fraction of material phases. By updating the definitions of $\mu$, $b_n$, and $c_n$ in the above three equations, the results from Willis's original work shall be identical to the re-derivation results.

\subsection{Numerical results of Green's function}
Using the re-derived Floquet number and Fourier coefficients for ensemble averages, this subsection plots the Green's function with two realizations together with its (weighted) ensemble average. Note that the numerical results only aim to quantify the influence of these algebraic differences on the Green's function. Specifically, the published results in Fig. 1 and Fig. 3 in \cite{Willis2009} are compared with those obtained from the revised expressions, 

\begin{figure}[h!]
    \centering
    \includegraphics[width=1\linewidth]{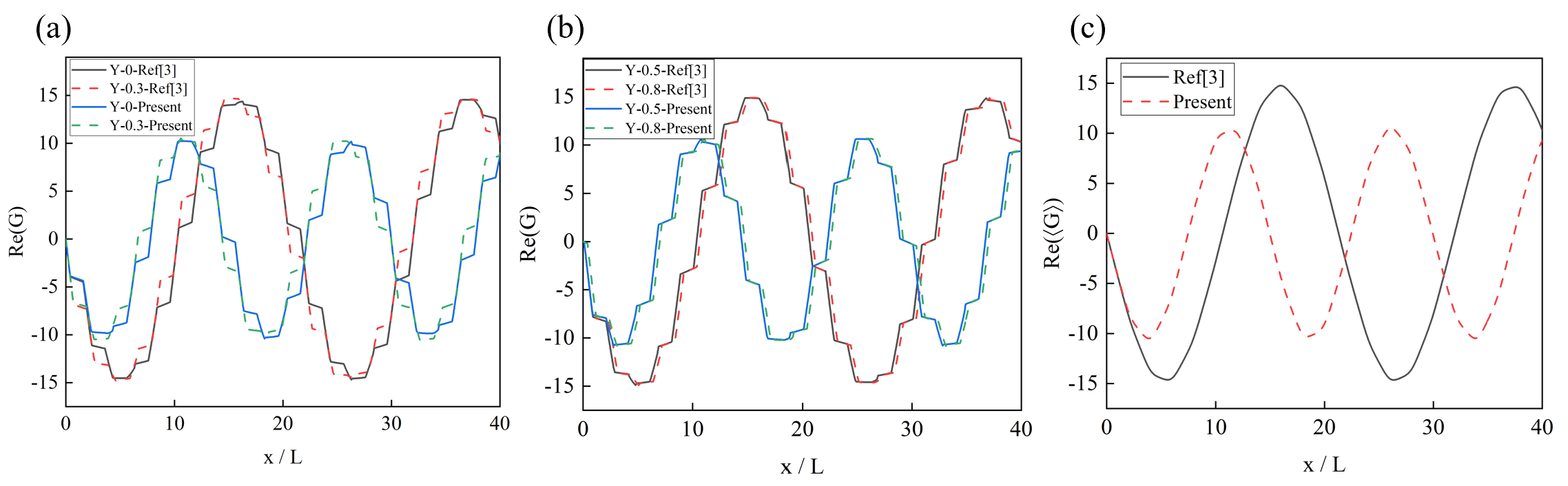}
    \caption{Variation of the real part of Green's function with realizations (a) $Y = 0, 0.3$; (b) $Y = 0.5, 0.8$; and (c) ensemble-averaged Green's function. The dimensionless coordinate $x / L \in [0, 40]$ at the frequency $L \omega / \sqrt{E_2 / \rho_2} = 0.1$, and the legends ``Ref [3]'' and ``Present'' refer to the published results in \cite{Willis2009} and the revised case, respectively.}
    \label{fig:Green_function}
\end{figure}
Figs. \ref{fig:Green_function} (a-c) follow Fig. 1 in \cite{Willis2009}, which plot the realization $Y= 0, 0.3, 0.5, 0.8$ and the ensemble-averaged Green's function ($w_y = 1$), when the normalized frequency $L \omega / \sqrt{E_2 / \rho_2} = 0.1$. Subsequently, Figs. \ref{fig:Green_function_3} (a-b) follows Fig. 3 in \cite{Willis2009}, which plots the weighted ensemble-averaged Green's function when the normalized frequency $L \omega / \sqrt{E_2 / \rho_2} = 0.1$ and $0.4$, respectively. Comparing the published and present results, noticeable changes in the amplitude, phase, and attenuation characteristics can be observed. These differences indicate that the re-derived formulae above have a non-negligible influence on the microstructure-specific Green's function. However, at this stage, comparing the Green's function alone does not indicate that it can provide physically consistent predictions of effective kernels, which will be further evaluated and compared in Section \ref{sec:example}. 
\begin{figure}[h!]
    \centering
    \includegraphics[width=0.8\linewidth]{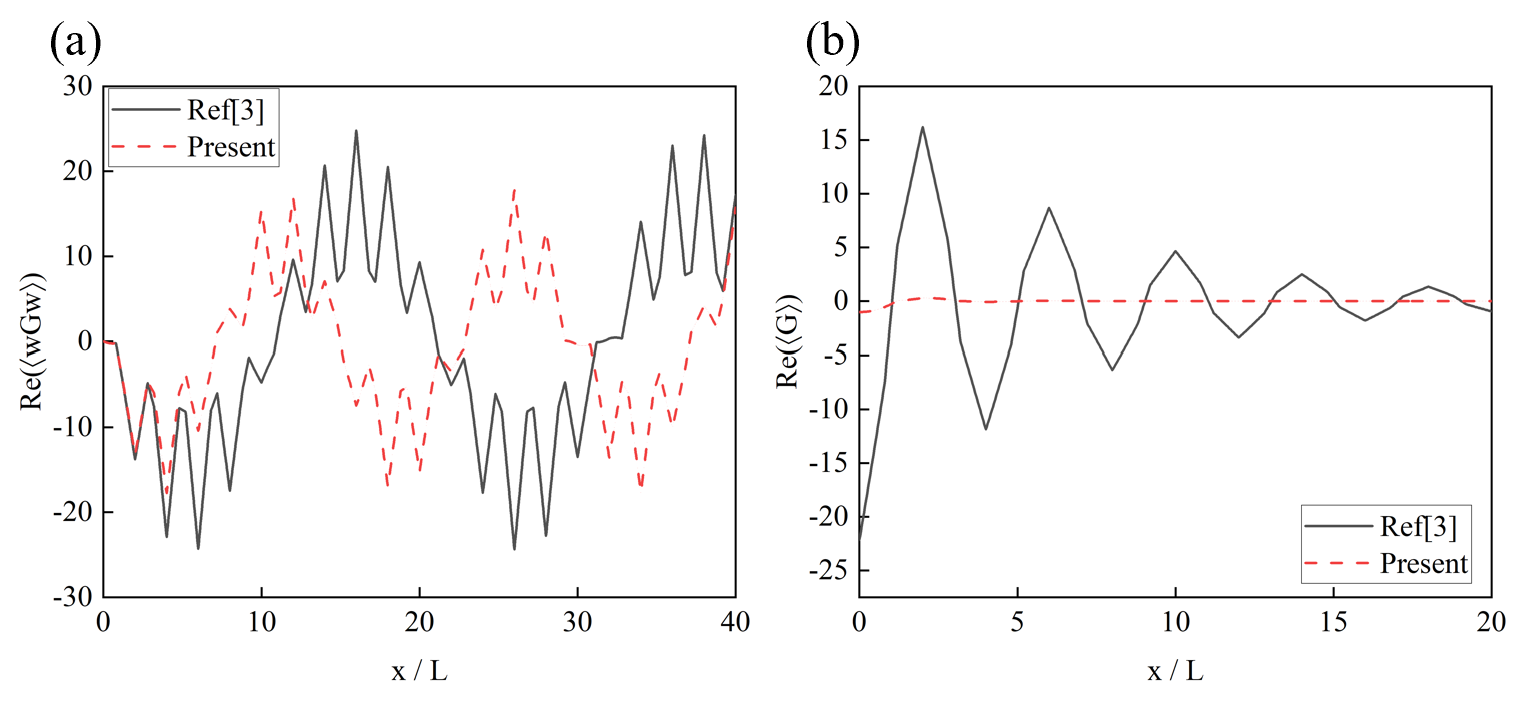}
    \caption{Variation of the real part of the weighted mean Green's function with two normalized frequencies (a) $L \omega / \sqrt{E_2 / \rho_2} = 0.1$; and (b) $L \omega / \sqrt{E_2 / \rho_2} = 0.4$. The legends ``Ref [3]'' and ``Present'' refer to the published results in \cite{Willis2009} and the revised case, respectively.}
    \label{fig:Green_function_3}
\end{figure}
\\
\section{Homogenization framework using the Eshelby's equivalent inclusion method}
Fu and Mura \cite{Fu1983} first proposed the EIM in elastodynamics. In analogy with EIM in elastostatics, the inhomogeneity is replaced by a matrix and two ``eigenstrains '' that simulate material mismatches in stiffness and mass density, respectively. Following the pioneering work \cite{Fu1983}, some subsequent works \cite{srivastava2012overall, Song2018, Wu2025} have introduced different forms of eigen-fields, including eigenstrain, eigenstress, eigen-momentum, and eigen-body-force. However, these eigen-fields were primarily auxiliary fields to reproduce the local responses of the composites, provided they are consistently solved through equivalent conditions. In the following, the present work employs eigenstrain and eigen-momentum as auxiliary fields to simulate the mismatches in stiffness and mass density, respectively. This section provides a general approach with the EIM to reproduce local responses and evaluate Willis effective kernels without deriving microstructure-specific Green's functions.

\subsection{Formulation of the Eshelby's equivalent inclusion method for local fields}
The Green's function for an infinite one-dimensional domain with a unit harmonic point force can be written in the space domain as, 
\begin{equation}
    G^H(x, x') = \frac{i}{2 E_c \gamma} \exp[i \gamma r], \quad \gamma = \omega \sqrt{\frac{\rho_c}{E_c}}, \quad r = |x - x'|
    \label{eq:gfunc}
\end{equation}
where $\rho_c$ and $E_c$ are the mass density and elastic modulus of the comparison medium, respectively; and $G^H$ refers to the elastodynamic Green's function of the comparison medium, which satisfies the governing equation $E_c G^H_{,xx}(x, x') + \rho_c \omega^2 G^H(x, x') = -\delta(x - x')$, which is similar to but different from Eq. \eqref{eq:green_gov}. The local responses and the resulting effective kernels are independent of the comparison medium, provided the eigen-fields are solved consistently. Here, the comparison medium is used for the subsequent derivations, but one may employ one material phase as the comparison medium to avoid deriving eigen-fields in that material phase and thus reduce the system's degrees of freedom. 

Since this section employs eigenstrain and eigen-momentum as two eigen-fields, they should be determined through the equivalent conditions as, 
\begin{equation}
\begin{split}
    \sigma_{Y}(x) &= E_c \left[ \varepsilon(x) - \varepsilon^e(x) - \varepsilon^*_{Y}(x) \right] = E(x) (u_{Y,x}(x) - \varepsilon^e), \\  
     p_Y(x) &= -i \omega \rho_c u_Y(x) - \eta_{Y}^*(x) = - i \omega \rho(x) u_{Y}(x)
\end{split}
\label{eq:equiv_cond}
\end{equation}
where $\varepsilon_{Y}^*(x)$ and $\eta_Y^*$ are eigenstrain and eigen-momentum to simulate the material mismatches in stiffness and density, respectively, in realization $Y$; and $\varepsilon^e(x)$ is a prescribed eigenstrain for the eigen-field homogenization framework in Section 5. Replacing stress and momentum in Eq. (\ref{eq:gov_eqn}), the governing equation can be written for one homogeneous comparison medium containing spatially varying eigen-fields, 
\begin{equation}
    E_c \left[ \varepsilon_{,x}(x) - \varepsilon_{,x}^e(x) - \varepsilon^*_{Y,x}(x) \right] + \omega^2 \rho_c u_{Y}(x) - i \omega \eta_Y^*(x) + f(x) = 0 
    \label{eq:gov_trans}
\end{equation}
Using the filtering effect of the Dirac delta function together with the Green's function, the displacement for any interior points can be derived as, 
\begin{small}
\begin{equation}
    \begin{split}
    u_Y(x) &= \int_{-L}^{L} \delta(x - x') u_Y(x') \thinspace dx' = -\int_{-L}^{L} \left[ E_c G_{,xx}^H(x, x') + \rho_c \omega^2 G^H(x, x') \right] \thinspace u_Y(x') \thinspace dx' \\ 
    & = \Big\{ - \left[ E_c G_{,x'}^H(x, L) u_{Y}(L) - E_c G^H_{,x'}(x, -L) u_Y(-L) \right] \\ & + E_c G^H(x, L) n(L) [u_{Y,x'}(L) - \varepsilon^e(L) - \varepsilon_{Y}^*(L) ] 
    \thinspace + E_c G^H(x, -L) n(-L) [u_{Y,x'}(-L)- \varepsilon^e(-L) - \varepsilon_{Y}^*(-L) ] \Big\} \\ &+ \int_{-L}^L -E_c G_{,x}^H(x, x') \left[ \varepsilon_Y^*(x') + \varepsilon^e(x') \right] \thinspace dx' 
    \thinspace + \int_{-L}^L -i \omega G^H(x, x') \eta_Y^*(x') \thinspace dx' + \int_{-L}^L G^H(x, x') f(x') \thinspace dx' 
    \end{split}
    \label{eq:disp_full}
\end{equation}
\end{small}
where $-E_c (\varepsilon^*_{Y,x'} + \varepsilon^e_{,x'}) - i \omega \eta^*_Y(x') + f(x')$ is regarded as the fictitious source. The above equation can be written as 
\begin{equation}
    u_Y(x)= u_Y^b(x) + u_Y^s(x) + u_Y^e(x) + u_Y^m(x) + u_Y^f(x)
\end{equation}
where the superscripts $b, s, e, m, f$ refer to disturbances by boundary responses, mismatch eigenstrain, prescribed eigenstrain, eigen-momentum, and prescribed body force, respectively, as 
\begin{equation*}
\begin{aligned}
    u_Y^b(x) &=  - \left[ E_c G_{,x'}^H(x, L) u_{Y}(L) - E_c G^H_{,x'}(x, -L) u_Y(-L) \right] + E_c G^H(x, L) n(L) [u_{Y,x'}(L) - \varepsilon^e(L) - \varepsilon_{Y}^*(L) ] \\   
    & \quad \thinspace + E_c G^H(x, -L) n(-L) [u_{Y,x'}(-L)- \varepsilon^e(-L) - \varepsilon_{Y}^*(-L) ] \\
    u_Y^s(x) &= -\int_{-L}^L E_c G_{,x}^H(x, x') \varepsilon_Y^*(x') \thinspace dx', \quad 
    u_Y^e(x) = -\int_{-L}^L E_c G_{,x}^H(x, x') \varepsilon^e(x') \thinspace dx' \\
    u_Y^m(x) &= \int_{-L}^L -i \omega G^H(x, x') \eta_Y^*(x') \thinspace dx', \quad
    u_Y^f(x) = \int_{-L}^L G^H(x, x') f(x') \thinspace dx'
\end{aligned}
\end{equation*}
Note that the traction, $t_Y = E_c \thinspace n \thinspace (u_{Y,x} - \varepsilon^e - \varepsilon^*_Y)$, contains the contribution from the prescribed eigenstrain and mismatch eigenstrain. For instance, when the realization $Y = 0$, only if the material phase 2 (dark gray) is selected as the comparison medium does the eigenstrain on the boundary become zero. Specifically, the eigen-fields can be determined using Eq. (\ref{eq:equiv_cond}), 
\begin{equation}
    \varepsilon_{Y}^* = \frac{E_c - E(x)}{E_c} [\varepsilon_Y(x) - \varepsilon^e(x)], \quad \eta_Y^*(x) = -i \omega [\rho_c - \rho(x)] u_Y(x) + M^*_{Y} \delta(x - \beta_Y)
    \label{eq:sol_eigen}
\end{equation}
where $M_Y^* = i \omega M u_{Y}(\beta_Y)$ simulates the momentum introduced by the concentrated mass $M$. As Eq. (\ref{eq:sol_eigen}) indicates, the eigen-fields are generally spatially varying. Therefore, this section discretizes the composite domain into $N$ line elements of length $2L / N$. When the line elements are relatively short, the eigen-fields can be assumed constant over each line element. Consequently, the piecewise constant eigen-fields can be represented in terms of Heaviside Theta functions ($\Theta(.)$) \cite{Michelitsch2003}, 
\begin{equation}
    \varepsilon_Y^*(x) = \sum_{I=1}^{N} \varepsilon_{Y}^{I*} \Theta \left[ \left( \frac{L}{N} \right)^2 - (x - x^{IC})^2 \right], \quad \eta_Y^*(x) = \sum_{I=1}^{N} \eta_Y^{I*} \Theta \left[ \left( \frac{L}{N} \right)^2 - (x - x^{IC})^2 \right] + M_Y^* \delta(x - \beta_Y)
    \label{eq:eigens}
\end{equation}
where $\varepsilon^{I*}_Y$ and $\eta_Y^{I*}$ refer to constant eigenstrain and eigen-momentum over the $\text{I}^\text{th}$ line element; and $\Theta [ \left( \frac{L}{N} \right)^2 - (x - x^{IC})^2 ] = \Theta[\frac{L}{N} - (x - x^{IC})] \thinspace \Theta[\frac{L}{N} + (x - x^{IC})]$, where $x^{IC}$ is the center of the $\text{I}^\text{th}$ line element. The disturbances by two eigen-fields and the body force can be evaluated through domain integrals of Green's function, 
\begin{equation}
\begin{aligned}
    u_Y^s(x) & = \sum_{I=1}^{N} \varepsilon_Y^{I*} \int_{x^{IC} - \frac{L}{N}}^{x^{IC} + \frac{L}{N}} -E_c G^H_{,x}(x, x') \thinspace  dx' = \sum_{I=1}^{N} S^I(x) \varepsilon_{Y}^{I*} \\
    u_Y^m(x) & = -i \omega \left( M_Y^* G^H(x, \beta_Y) + \sum_{I=1}^{N} \eta_Y^{I*} \int_{x^{IC} - \frac{L}{N}}^{x^{IC} + \frac{L}{N}} G^H(x, x') \thinspace  dx' \right) = -i \omega \left( M_Y^* G^H(x, \beta_Y) + \sum_{I=1}^{N} L^I(x) \eta_{Y}^{I*} \right)
\end{aligned}
\end{equation}
where $L^I$ and $S^I = -E_c L^I_{,x}$ are similar to Eshelby's tensors for eigen-momentum and eigenstrain, respectively, in 2D or 3D problems although they are scalar for the 1D case. Without loss of generality, let $a$ and $b$ denote the starting and ending points of the line element; the closed-form Eshelby's tensor can be obtained \cite{Wu2026_rspa}, 
\begin{equation}
    L(x) = \int_{a}^{b} \frac{i}{2 E_c \gamma} \exp [i \gamma r] \thinspace dx' = \frac{-1}{2 E_c \gamma^2} \begin{cases} 
    \exp[i \gamma (a - x)] - \exp[i \gamma (b - x)] & x \in (-\infty, a) \\ 
    2 - \exp[i \gamma (x - a)] - \exp[i \gamma (b - x)] & x \in [a, b] \\
    \exp[i \gamma (x - b)] - \exp[i \gamma (x - a)] & x \in (b, +\infty)
    \end{cases}
\end{equation}
and its first order derivative, 
\begin{equation}
    L_{,x}(x) = \frac{d}{dx} \int_{a}^{b} \frac{i}{2 E_c \gamma} \exp [i \gamma r] \thinspace dx' = \frac{-i}{2 E_c \gamma} \begin{cases} 
    \exp[i \gamma (b-x)]  - \exp[i \gamma (a-x)] & x \in (-\infty, a) \\ 
    \exp[i \gamma (b - x)] - \exp[i \gamma (x - a)]  & x \in [a, b] \\
    \exp[i \gamma (x - b)] - \exp[i \gamma (x - a)]  & x \in (b, +\infty)
    \end{cases}
\end{equation}

Hence, a global linear equation system can be constructed, including $ 2N + 1$ unknown eigen-fields and 2 unknown boundary responses. Note that the global equation system enforces the boundary conditions and equivalent conditions together as follows: 
\begin{equation}
\begin{split}
    &\begin{bmatrix} {\mathcal{H}} & \ldots &  i \omega L^I & i \omega G^H & -S^I & \ldots \\ \vdots & \vdots & \vdots & \vdots &\vdots & \vdots \\ -i \omega \Delta \rho {\mathcal{H}} & \ldots & \omega^2 \Delta \rho L^{I} - 1 & \omega^2 \Delta \rho G^{H} & -i \omega \Delta \rho S^{I} & \ldots \\ 
    -i \omega M {\mathcal{H}} & \ldots & \omega^2 M L^I & \omega^2 M G^H - 1 & -i \omega M S^I & \ldots \\ 
    \Delta E {\mathcal{H}}' & \ldots & i \omega \Delta E L^{I'} & i \omega \Delta E G^{H'} & \Delta E S^{I'} - E_c & \ldots \\ 
    \vdots & \vdots & \vdots & \vdots & \vdots & \vdots
    \end{bmatrix} \begin{bmatrix} u_Y \\ \vdots \\ \eta_Y^{I*} \\ M_Y^* \\ \varepsilon_Y^{I*} \\ \vdots \end{bmatrix} \\ & = \begin{bmatrix} {\mathcal{G}} \\ \vdots \\ -i \omega \Delta \rho {\mathcal{G}} \\ -i \omega M {\mathcal{G}} \\  \Delta E {\mathcal{G}}' \\ \vdots \end{bmatrix} \begin{bmatrix} t_Y \end{bmatrix} + \begin{bmatrix} L & S \\ \vdots & \vdots \\ -i\omega \Delta \rho L & -i \omega \Delta \rho S \\ -i \omega M L & - i \omega M S \\ \Delta E L' & \Delta E (S' - 1) \\ \vdots & \vdots \end{bmatrix} \begin{bmatrix} f \\ \varepsilon^e \end{bmatrix}
\end{split}
    \label{eq:global}
\end{equation}
where $\mathcal{H} = E_c G^H_{,x}(x, \pm L)$ and $\mathcal{G} = G^H(x, \pm L)$ evaluates contributions by boundary responses at two ends; $u_Y$ and $t_Y$ are boundary displacement and traction with the realization $Y$; the superscript $(.)'$ denotes the partial differentiation with respect to $x$. In addition, $\Delta \rho = \rho_c - \rho(x)$ and $\Delta E = E_c - E(x)$ represent the material mismatches in mass density and elastic modulus, respectively. 

As previously mentioned in Section 2, the composite system is subjected to the Bloch-form boundary condition. In Eq. (\ref{eq:bloch}), the displacement and traction at the two ends are linearly related. Without loss of generality, this section employs the boundary responses at the starting (left) point as the unknowns; the global linear equation system can be rearranged formally as, 
\begin{equation}
    \bm{H}_Y \bm{\mathcal{S}}_Y = \bm{\mathcal{M}}_Y \bm{D}
\end{equation}
where $\bm{H}_Y$ refers to the rearranged matrix of coefficients; $\bm{\mathcal{S}}_Y$ is the response vector, which includes the boundary responses of the starting point and all eigen-fields; and $\bm{D}$ stands for the matrix of driving sources, including body forces and prescribed eigenstrains. Specifically, they can be written as, 
\begin{equation}
\begin{split}
    &\begin{bmatrix} {\mathcal{H}}^L + \lambda \mathcal{H}^R & -\mathcal{G}^L + \lambda \mathcal{G}^R & \ldots &  i \omega L^I & i \omega G^H & -S^I & \ldots \\ \vdots & \vdots & \vdots & \vdots &\vdots & \vdots & \vdots \\ -i \omega \Delta \rho ({\mathcal{H}}^L + \lambda \mathcal{H}^R) & i \omega \Delta \rho (\mathcal{G}^L - \lambda \mathcal{G}^R) & \ldots & \omega^2 \Delta \rho L^{I} - 1 & \omega^2 \Delta \rho G^{H} & -i \omega \Delta \rho S^{I} & \ldots \\ 
    -i \omega M ({\mathcal{H}}^L + \lambda \mathcal{H}^R) & i \omega M (\mathcal{G}^L - \lambda \mathcal{G}^R) & \ldots & \omega^2 M L^I & \omega^2 M G^H - 1 & -i \omega M S^I & \ldots \\ 
    \Delta E ({\mathcal{H}}^{L'} + \lambda \mathcal{H}^{R'}) & -\Delta E (\mathcal{G}^{L'} - \lambda \mathcal{G}^{R'}) & \ldots & i \omega \Delta E L^{I'} & i \omega \Delta E G^{H'} & \Delta E S^{I'} - E_c & \ldots \\ 
    \vdots & \vdots & \vdots & \vdots & \vdots & \vdots & \vdots
    \end{bmatrix} \begin{bmatrix} u_Y^L \\ t_Y^L \\ \vdots \\ \eta_Y^{I*} \\ M_Y^* \\ \varepsilon_Y^{I*} \\ \vdots \end{bmatrix} \\
    & = \begin{bmatrix} L & S \\ \vdots & \vdots \\ -i\omega \Delta \rho L & -i \omega \Delta \rho S \\ -i \omega M L & - i \omega M S \\ \Delta E L' & \Delta E (S' - 1) \\ \vdots & \vdots \end{bmatrix} \begin{bmatrix} f \\ \varepsilon^e \end{bmatrix}
\end{split}
    \label{eq:global_re}
\end{equation}
where $\lambda = \exp[2 i \zeta L]$. Since the prescribed body force $\bm{f}$ and the eigenstrain $\varepsilon^e$ are given, the response vector $\bm{\mathcal{S}}_Y$ can be determined as $\bm{\mathcal{S}}_Y = \bm{\mathcal{H}}_Y^{-1} \bm{\mathcal{M}}_Y \bm{D}$. Once the response vector is determined, the displacement and its gradient can be recovered through Eq. (\ref{eq:disp_full}).

\subsection{Extraction of effective properties from the explicit solution}
Effective properties characterize the composite's macroscopic performance by relating macroscopic stress and momentum to strain and velocity. Hence, it is natural to extract effective properties from field averages using explicit solutions. For instance, the composite system can be excited by body forces, and the field averages can be expressed as functions of body forces. By eliminating the body forces, the effective constitutive relations can be subsequently determined, namely the source-driven homogenization framework. 

As mentioned in Section 2, the conventional source-driven homogenization framework can lead to nonunique effective properties because the de-phased averaged displacement, velocity, and strain are kinematically constrained, leading to an admissible representation of the effective constitutive relation \cite{Willis2009}. Willis \cite{Willis2011} subsequently resolved this nonuniqueness issue by introducing additional residual fields. For a 1D laminate, either a prescribed eigenstrain or a prescribed body force can be introduced with the conventional body-force excitation to generate two independent macroscopic states, from which the complete and unique Willis constitutive relations can be determined. In the present work, we extract the effective properties with the eigen-field and source-driven framework of Willis \cite{Willis2011}, in which an additional eigenstrain is prescribed to construct the independent macroscopic state.

Before extracting effective properties, this subsection evaluates the de-phased averages using the equivalent inclusion method. Using Eq. (\ref{eq:disp_full}) and its partial derivative with respect to $x$, the velocity $-i \omega u_Y$, strain $\varepsilon_Y$, momentum $p_Y$, and stress $\sigma$ of the sampling point $x_p$ can be formally written as, 
\begin{equation}
    \begin{bmatrix} \vdots \\ \varepsilon_Y(x_p) \\ -i \omega u_Y(x_p) \\ \vdots \end{bmatrix}  = \bm{\mathcal{A}}_Y \bm{\mathcal{S}}_Y + \bm{\mathcal{T}}_Y^A \bm{D}, \quad \begin{bmatrix} \vdots \\ \sigma_{Y}(x_p) \\ p_Y(x_p) \\ \vdots \end{bmatrix}  = \bm{\mathcal{B}}_Y \bm{\mathcal{S}}_Y + \bm{\mathcal{T}}_Y^B \bm{D}
    \label{eq:post_strain_v}
\end{equation}
where these coefficient matrices are, 
\begin{equation}
    \begin{aligned}
    \bm{\mathcal{A}}_Y &= \exp[-i \zeta x_p] \begin{bmatrix} 
    \vdots & \vdots & \vdots & \vdots & \vdots & \vdots & \vdots \\
                    -\mathcal{H}' \thinspace w_Y & \mathcal{G}'\thinspace w_Y & \ldots & -i \omega L^{I\prime}\thinspace w_Y & -i \omega G^{H\prime}\thinspace w_Y & S^{I\prime}\thinspace w_Y  & \ldots \\ 
                    i \omega \mathcal{H}\thinspace w_Y & -i \omega \mathcal{G}\thinspace w_Y & \ldots & \omega^2 L^I \thinspace w_Y & \omega^2 G^H \thinspace w_Y &  -i \omega S^I \thinspace w_Y  & \ldots \\ 
                    \vdots & \vdots & \vdots & \vdots & \vdots & \vdots & \vdots 
    \end{bmatrix} \\ 
    \bm{\mathcal{B}}_Y &= \exp[-i \zeta x_p] \begin{bmatrix} 
    \vdots & \vdots & \vdots & \vdots & \vdots & \vdots & \vdots \\
                    - E_Y \mathcal{H}'  & E_Y \mathcal{G}'  & \ldots & -i \omega E_YL^{I\prime} & -i \omega E_Y G^{H\prime} & E_Y S^{I\prime}  & \ldots \\ 
                    i \omega \rho_Y \mathcal{H} & -i \omega \rho_Y \mathcal{G} & \ldots & \omega^2 \rho_Y L^I  & \omega^2 \rho_Y G^H  &  -i \omega \rho_Y S^I   & \ldots \\ 
                    \vdots & \vdots & \vdots & \vdots & \vdots & \vdots & \vdots 
    \end{bmatrix} \\
    \bm{\mathcal{T}}_Y^A & =\exp[-i \zeta x_p] \begin{bmatrix} \vdots & \vdots \\ L' \thinspace w_Y & S' \thinspace w_Y \\ -i \omega L \thinspace w_Y & -i \omega S \thinspace w_Y \\ \vdots & \vdots\end{bmatrix}, \qquad \bm{\mathcal{T}}_Y^B = \exp[-i \zeta x_p] \begin{bmatrix} \vdots & \vdots \\ E_Y L'  & E_Y (S' - 1) \\ -i \omega \rho_Y L  & -i \omega \rho_Y S  \\ \vdots & \vdots\end{bmatrix}
    \end{aligned}
    \label{eq:avg_eqn}
\end{equation}
where a de-phase factor of $\exp[-i \zeta x_p]$ is applied to all matrices; and $w_Y$ is the weight function. Note that the weight function is only applied for averaged strain and velocity. Furthermore, $\bm{\mathcal{A}}_Y$ and $\bm{\mathcal{B}}_Y$ are two coefficient matrices for strain, velocity, and stress, momentum, respectively, and similarly $\bm{\mathcal{T}}_Y^A$, $\bm{\mathcal{T}}_Y^B$ evaluate the contributions by the body forces. Because the elastic fields at all sampling points are linearly dependent on the response vector and body force, the de-phased averaged fields can be obtained through $NP$ sampling points, 
\begin{equation}
    \begin{aligned}
    \begin{bmatrix} \langle \varepsilon \rangle_Y \\ -i \omega \langle u \rangle_Y \end{bmatrix} = \frac{1}{NP} \sum_{i=1}^{NP} \left[ (\bm{\mathcal{A}}_Y)_{ip} (\bm{\mathcal{S}}_Y)_p + (\bm{\mathcal{T}}^A_Y)_{ip} (\bm{D})_p \right] = \left( \overline{\mathcal{A}_Y} \bm{H}_Y^{-1} \bm{\mathcal{M}}_Y + \overline{\bm{\mathcal{T}}^A_Y} \right) \bm{D} \\ 
    \begin{bmatrix} \langle \sigma \rangle_Y \\ \langle P \rangle_Y \end{bmatrix} = \frac{1}{NP} \sum_{i=1}^{NP} \left[ (\bm{\mathcal{B}}_Y)_{ip} (\bm{\mathcal{S}}_Y)_p + (\bm{\mathcal{T}}^B_Y)_{ip} (\bm{D})_p \right] + \begin{bmatrix} 0 \\ \frac{-i \omega M }{2L} \exp[-i \zeta \beta_Y] u_Y(\beta_Y) \end{bmatrix} = \left( \overline{\mathcal{B}_Y} \bm{H}_Y^{-1} \bm{\mathcal{M}}_Y + \overline{\bm{\mathcal{T}}^B_Y} \right) \bm{D}
    \end{aligned}
    \label{eq:dephase_average}
\end{equation}
where the term $\langle P \rangle_Y$ has involved the contribution from the concentrated mass. Now the de-phased averaged strain, velocity, stress, and momentum are linearly related to the driven sources $\bm{D}$. 

\subsubsection{Evaluate one admissible representative effective constitutive relation}
An admissible representative of the effective kernels is selected by requiring all components to be even functions of the Bloch wavenumber $\zeta$ \cite{Willis2009}. With the additional constraint, the responses associated with two opposite Bloch wavenumbers can be combined to determine the effective kernels. Note that only the body force is required in this procedure, while the prescribed eigenstrain is set to zero. Two macroscopic states provide 
\begin{equation}
    \langle \sigma \rangle_Y(\zeta) = -i \omega \tilde{S}^\text{eff} \langle u \rangle_Y + i \zeta \tilde{E}^\text{eff} \langle u \rangle_Y, \quad \langle p \rangle_Y(\zeta) = -i \omega \tilde{\rho}^\text{eff} \langle u \rangle_Y + i \zeta \tilde{S}^{\dagger, \text{eff}} \langle u \rangle_Y
\end{equation}
By defining 
\begin{equation}
\begin{aligned}
    I_1^Y(\zeta) = \frac{\langle \sigma \rangle_Y (\zeta)}{\langle u \rangle_Y (\zeta)} = - i \omega \tilde{S}^\text{eff} + i \zeta \tilde{E}^\text{eff}, \quad I_1^Y(-\zeta) = \frac{\langle \sigma \rangle_Y (-\zeta)}{\langle u \rangle_Y (-\zeta)} = - i \omega \tilde{S}^\text{eff} - i \zeta \tilde{E}^\text{eff} \\ 
    I_2^Y(\zeta) = \frac{\langle p \rangle_Y (\zeta)}{\langle u \rangle_Y (\zeta)} = -i \omega \tilde{\rho}^\text{eff} + i \zeta \tilde{S}^{\dagger, \text{eff}}, \quad I_2^Y(-\zeta) = \frac{\langle p \rangle_Y (-\zeta)}{\langle u \rangle_Y (-\zeta)} = -i \omega \tilde{\rho}^\text{eff} - i \zeta \tilde{S}^{\dagger, \text{eff}}
\end{aligned}
\end{equation}
the effective properties can be determined as, 
\begin{equation}
\begin{aligned}
    \tilde{E}^\text{eff} & = \frac{I_1^Y(\zeta)-I_1^Y(-\zeta)}{2i\zeta}, \quad \tilde{S}^\text{eff} = \frac{i \left[I_1^Y(\zeta)+I_1^Y(-\zeta)\right]}{2\omega} \\ \tilde{S}^{\dagger,\text{eff}}&=\frac{I_2^Y(\zeta)-I_2^Y(-\zeta)} {2i\zeta}, \quad \tilde{\rho}^\text{eff}=\frac{i \left[I_2^Y(\zeta)+I_2^Y(-\zeta)\right]}{2\omega}
\end{aligned}
\label{eq:effec_non}
\end{equation}
Note that when $\zeta \to 0^+$, the effective kernels in Eq. (\ref{eq:effec_non}) should be interpreted as the long-wavelength limit. For instance, Section 5.1 set $\zeta = 10^{-4}$ to approximate the long-wavelength limit. 

\subsubsection{Evaluate the unique representative effective constitutive relation}
In addition to the body force $f$, we can use an additional prescribed eigenstrain $\varepsilon^e$ to provide the second macroscopic state. Using the superposition theorem, the eigen-field approach can be conducted in two steps: (i) evaluate the averaged field quantities under the body-force excitation by setting $\varepsilon^e = 0$, which is denoted by the superscript $b$; and (ii) evaluate the averaged field quantities under the prescribed eigenstrain excitation by setting $f = 0$, which is denoted by the superscript $e$. Hence, the effective kernels can be determined from the two macroscopic states as, 
\begin{equation}
    \begin{bmatrix} \tilde{E}^\text{eff} & \tilde{S}^\text{eff} \\ \tilde{S}^{\dagger, \text{eff}} & \tilde{\rho}^\text{eff} \end{bmatrix} =  \begin{bmatrix} \langle \sigma \rangle^{b}_Y & \langle \sigma \rangle_Y^e \\ \langle p \rangle_Y^b & \langle p \rangle^e_Y \end{bmatrix} \begin{bmatrix} \langle \varepsilon \rangle_Y^b & \langle \varepsilon - \varepsilon^e \rangle_Y^e \\ -i \omega \langle u \rangle_Y^b & -i \omega \langle u \rangle_Y^e \end{bmatrix}^{-1}
    \label{eq:extract_residual}
\end{equation}

The additional prescribed eigenstrain excitation removes the kinematic dependence between the de-phased averaged strain and velocity. For instance, if $w_Y = 1$, the body force excitation case satisfies $\langle \varepsilon \rangle_Y = i \zeta \langle u \rangle_Y$, while the prescribed eigenstrain excitation case becomes $\langle \varepsilon - \varepsilon^e \rangle_Y^e = i \zeta \langle u \rangle_Y^e - \langle \varepsilon^e \rangle$. The additional prescribed eigenstrain breaks the original kinematic constraint and provides a second independent macroscopic state; thus, the Willis effective kernels can be uniquely determined.

\section{Re-evaluated effective properties} 
\label{sec:example}
The preceding sections re-derived the microstructure-specific Green's function used in the exact homogenization method and proposed an EIM-based homogenization formulation to evaluate effective kernels. This section revisits the numerical examples reported in \cite{Willis2009, Willis2012} and examines how the differences identified above change the effective kernels. We compare the published results with the present re-derived exact-homogenization results and EIM-based homogenization results. Section 5.1 considers the admissible representative defined in \cite{Willis2009}, and Section 5.2 shows the unique effective constitutive relation using residual fields as in \cite{Willis2012}. Note that the driving sources, $f$ and $\varepsilon^e$, are presented in the Bloch-form, which should include the phase part $\exp[i \zeta x]$ and the periodic part. When the weight function $w_Y \neq 1$ is not uniform, the body force must be weighted accordingly, while the prescribed eigenstrain should not be weighted. In the following case studies, the exact homogenization method utilizes $201$ Fourier series (truncated at $|m| \leq 100$) to assure the convergent results. The EIM discretizes the laminated composite into $200$ line elements, and the de-phased averages are evaluated with $NP = 2,000$ sampling points, which are sufficient for convergence.

\subsection{Re-evaluation of the admissible effective kernels}
This subsection revisits Figs. 4--6 in \cite{Willis2009}. The laminated composite is subjected to the Bloch-form boundary condition with the macroscopic wavenumber $\zeta \to 0^+$. The material properties are defined as: (i) $E_1 = 0.05, \rho_1 = 0.5$; (ii) $E_2 = 1, \rho_2 = 0.5$; and (iii) the concentrated mass $M = 1$, so that the effective mass density at the homogenization limit $\langle \rho \rangle = 1$. 
\begin{figure}[h!]
    \centering
    \includegraphics[width=1\linewidth]{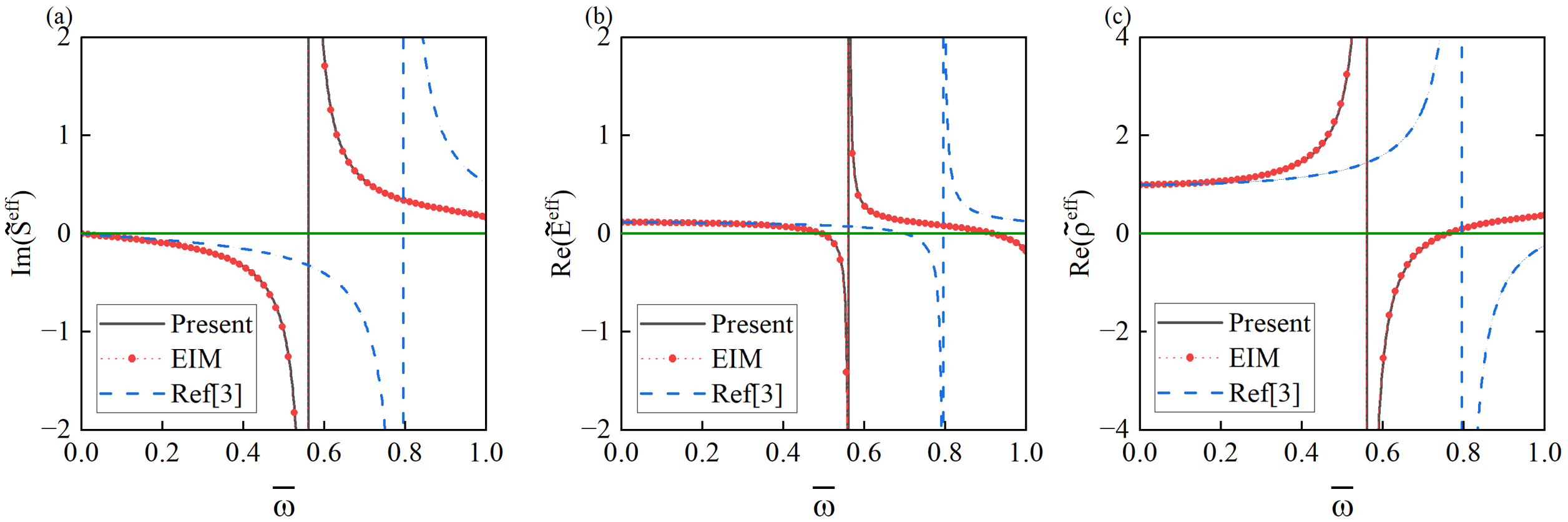}
    \caption{Comparison of the effective kernels obtained from \cite{Willis2009} (Ref [3]), the present re-derived exact homogenization method (Present), and the EIM homogenization method (EIM) for the uniform weight function $w_Y= 1$. (a) $Im(\tilde{S}^\text{eff})$; (b) $Re(\tilde{E}^\text{eff})$; and (c) $Re(\tilde{\rho}^\text{eff})$. The normalized frequency $\overline{\omega} = L \omega / \sqrt{E_2 / \rho_2}$.}
    \label{fig:2009_unweighted}
\end{figure}

Fig. \ref{fig:2009_unweighted} compares the effective kernels obtained from \cite{Willis2009}, the present re-derived exact homogenization method, and the EIM homogenization framework for the uniform weight function $w_Y = 1$. For brevity, the numerical curves reported in \cite{Willis2009}, the re-evaluated exact homogenization results, and the EIM results are denoted as ``Ref [3]'', ``Present'', and ``EIM''.  The exact homogenization method and EIM agree well over the entire frequency range for all three effective kernels. The published numerical results, however, differ noticeably from those two independently obtained results. Specifically, the pole-like responses predicted by the original results exist around $\overline{\omega} \approx 0.795$, while the present re-derived exact homogenization method and EIM both provide around $0.562$. The discrepancies also appear in the amplitude and phase delay of coupling, effective mass density, and elastic modulus. Therefore, these comparisons indicate that algebraic differences identified above can lead to corresponding shifts in the frequency-dependent effective kernels. 

\begin{figure}[h!]
    \centering
    \includegraphics[width=1\linewidth]{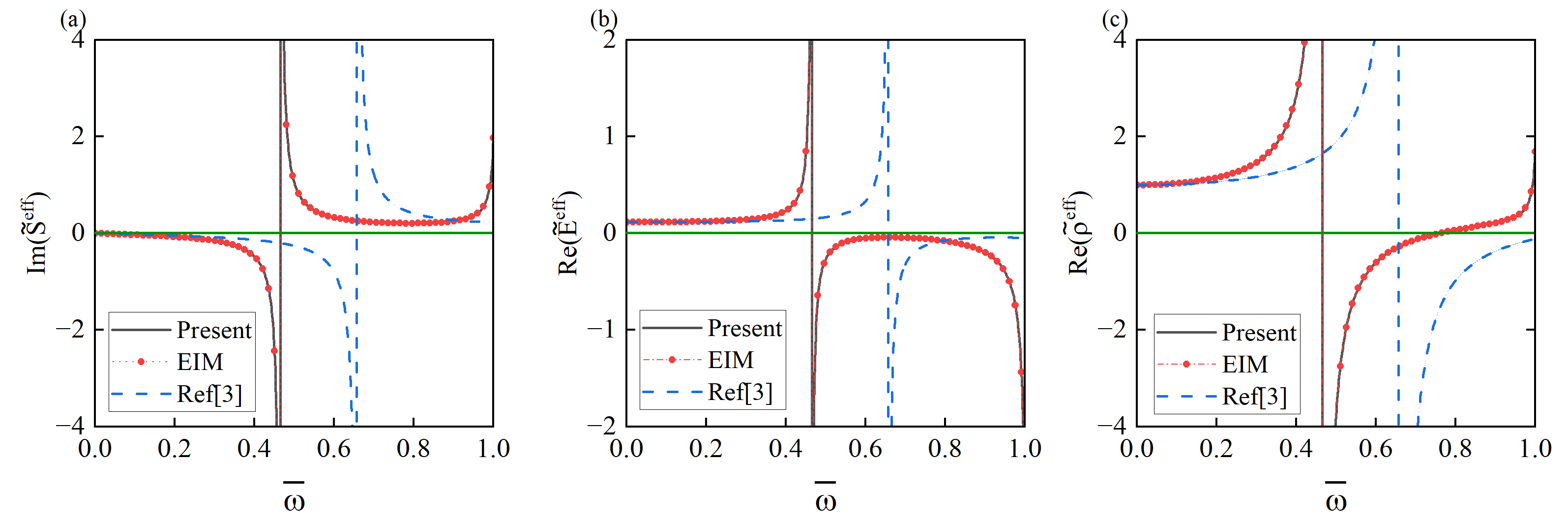}
    \caption{Comparison of the effective kernels obtained from \cite{Willis2009} (Ref [3]), the present re-derived exact homogenization method (Present), and the EIM homogenization method (EIM) for the weight function $w_Y= 1/c_2$ in the second material phase. (a) $Im(\tilde{S}^\text{eff})$; (b) $Re(\tilde{E}^\text{eff})$; and (c) $Re(\tilde{\rho}^\text{eff})$. The normalized frequency $\overline{\omega} = L \omega / \sqrt{E_2 / \rho_2}$.} 
    \label{fig:2009_weighted}
\end{figure}

Subsequently, Fig. \ref{fig:2009_weighted} compares the effective kernels when the weight function $w_Y = 1 / c_{2}$ is used in the second material phase. As shown in Fig. \ref{fig:2009_weighted}, the present re-derived exact homogenization solution agrees with the present EIM for all three effective kernels. At the same time, discrepancies appear between the original and present numerical methods. Similar to Fig. \ref{fig:2009_unweighted}, the original numerical results exhibit a delay of the pole-like responses. Therefore, the agreement between the exact homogenization method and the EIM is not restricted to the uniform-weight case, providing an independent cross-validation of the present re-evaluated results. 

\subsection{Re-evaluation of the unique effective kernels}
The unique effective kernels reported in Figs. 1--3 in \cite{Willis2012} are further revisited using the eigen-field formulae. The numerical case studies considered in \cite{Willis2012} are recalculated using the present re-derived exact homogenization method and the present EIM formulae. The comparison intends to check whether the published results can be produced consistently by two methods. The laminated composite is subjected to the Bloch-form boundary condition with the macroscopic wavenumber $\zeta = 0.1$ and $2$. The material properties are defined as: (i) $E_1 = 0.05 - 0.01i, \rho_1 = 0.5$; (ii) $E_2 = 1 - 0.1i, \rho_2 = 0.5$; and (iii) the concentrated mass $M = 1$, so that the effective mass density at the homogenization limit $\langle \rho \rangle = 1$. 

\begin{figure}
    \centering
    \includegraphics[width=0.67\linewidth]{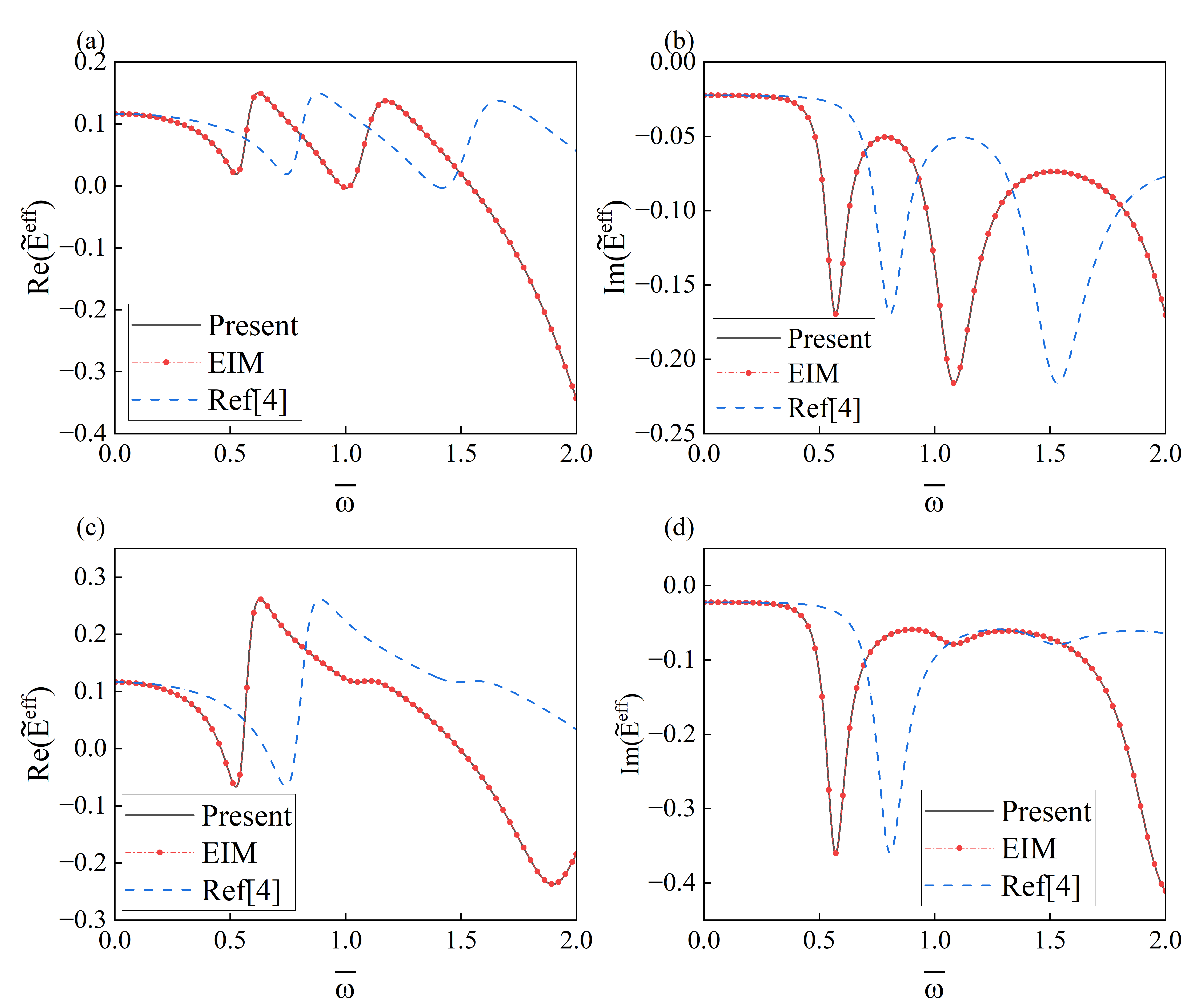}
    \caption{Comparison of the effective kernels obtained from \cite{Willis2012} (Ref [4]), the present re-derived exact homogenization method (Present), and the EIM homogenization method (EIM) with residual fields for the weight function $w_Y= 1$. 
    (a) $Re(\tilde{E}^\text{eff})$ and (b) $Im(\tilde{E}^\text{eff})$ with the admissible effective kernels; and (c) $Re(\tilde{E}^\text{eff})$ and (d) $Im(\tilde{E}^\text{eff})$ with the unique effective kernels. The normalized frequency $\overline{\omega} = L \omega / \sqrt{Re(E_2) / \rho_2}$, and the macroscopic wavenumber $\zeta = 0.1$.}
    \label{fig:unweight_01}
\end{figure}
\begin{figure}
    \centering
    \includegraphics[width=0.67\linewidth]{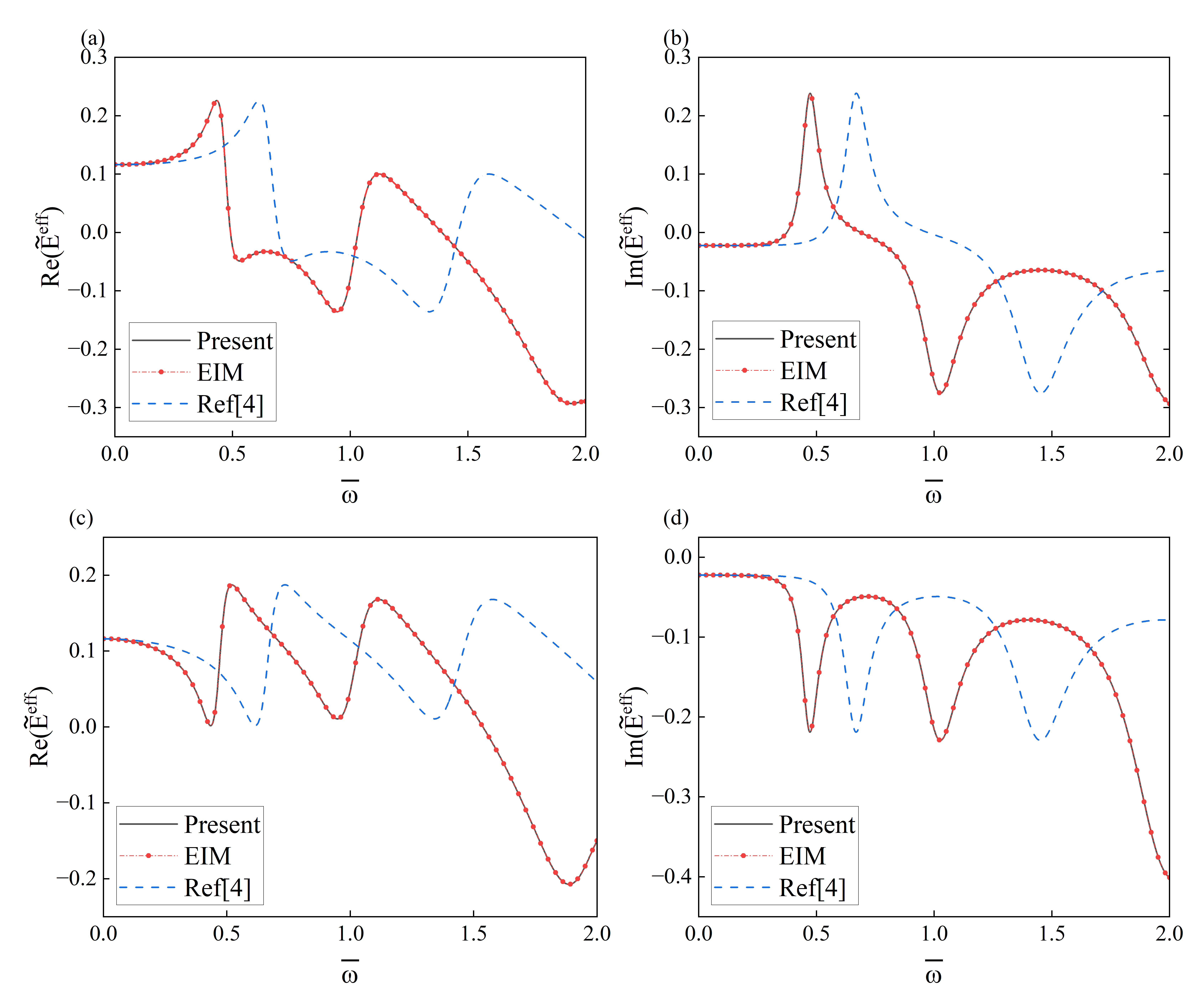}
    \caption{Comparison of the effective kernels obtained from \cite{Willis2012} (Ref [4]), the present re-derived exact homogenization method (Present), and the EIM homogenization method (EIM) with residual fields for the weight function $w_Y= 1/c_{2}$ in the second material phase. 
    (a) $Re(\tilde{E}^\text{eff})$ and (b) $Im(\tilde{E}^\text{eff})$ with the admissible effective kernels; and (c) $Re(\tilde{E}^\text{eff})$ and (d) $Im(\tilde{E}^\text{eff})$ with the unique effective kernels. The normalized frequency $\overline{\omega} = L \omega / \sqrt{Re(E_2) / \rho_2}$, and the macroscopic wavenumber $\zeta = 0.1$.}
    \label{fig:weight_01}
\end{figure}

Figs. \ref{fig:unweight_01} and \ref{fig:weight_01} compare the effective kernels reported in \cite{Willis2012} with those obtained from the present re-derived exact homogenization method and the EIM homogenization method with residual fields. Specifically, Figs. \ref{fig:unweight_01}(a) and (b) reproduce the real and imaginary parts shown in Fig. 1(a) of \cite{Willis2012}, whereas Figs. \ref{fig:unweight_01}(c) and (d) correspond to those in Fig. 1(c). Similarly, Figs. \ref{fig:weight_01}(a) and (b) correspond to the real and imaginary parts in Fig. 1(b), while Figs. \ref{fig:weight_01}(c) and (d) correspond to those in Fig. 1(d). Additional comparisons for the larger macroscopic wavenumber $\zeta = 2$, and the remaining Willis coupling, effective density kernels are provided in Appendix A.

Across all five comparisons, the present re-derived exact homogenization method and the EIM with residual fields show good agreement over the entire frequency range for both real and imaginary parts. In contrast, the published numerical curves are not recovered by either formulation, when the material properties and normalized frequencies written in \cite{Willis2012} are used. Moreover, systematic shifts appear in frequency-dependent features, such as phase delays. Therefore, these comparisons indicate that the differences identified in the microstructure-specific Green's function affect the resulting effective kernels and change their frequency-dependent performance. However, this discrepancy concerns only the published numerical results, and the underlying residual-field homogenization framework remains unaffected.

\section{Conclusions}
This paper revisited Willis' exact homogenization of one-dimensional periodically laminated composites and re-examined several differences in the analytical expressions and numerical results. We first re-derived the microstructure-specific Green's function and obtained a Floquet-number relation with a difference in the term associated with the concentrated mass, along with several differences in the Fourier coefficients. These revisions lead to changes in the evaluated magnitude, phase, and attenuation performance, which further affect the resulting effective kernels. Subsequently, we develop an equivalent inclusion-based homogenization framework using the infinite-domain Green's function of a homogeneous comparison medium. It shows that the effective kernels predicted by the present re-derived exact homogenization method and the EIM agree well. At the same time, the numerical curves reported in \cite{Willis2009, Willis2012} do not coincide with the results obtained under the material properties and normalized frequencies written in the text. Therefore, the present results provide independently cross-validated benchmarks for one-dimensional periodic laminated composites, and the EIM framework can be further extended for multiphysical and multidimensional Willis-type homogenization with more complex microstructures.

\section*{Data reproduction statement}
The source code (``C++'') to reproduce the present numerical results is provided in the Supplemental Materials. Detailed instructions for compilation, parameter settings, and reproduction of the exact homogenization method and equivalent inclusion method are included in the supplemental\_material.pdf.

\section*{Declaration of Competing Interest}
The authors declare that there is no conflict of interest.

\section*{CRediT author statement}
\textbf{Chunlin Wu}: Conceptualization, Methodology, Data Curation, Validation, Writing-Original Draft, Visualization; 
\textbf{Huiming Yin}: Conceptualization, Writing-Review \& Editing.

\section*{Acknowledgement}
The authors thank Drs. Gal Shmuel and John Willis for introducing this topics to us and their recommendation of publishing this communication note. CW's work is supported by National Natural Science Foundation of China Grant No. 12302086. HY's work was sponsored by the US Department of Agriculture NIFA \#2021-67021-34201 and the National Science Foundation (NSF) (IIP \#1738802 and IIP \#1941244). These supports are gratefully acknowledged.

\appendix 

\setcounter{figure}{0}
\renewcommand{\thefigure}{A.\arabic{figure}}

\section{Additional comparisons for different weight functions, wavenumbers, and effective kernels}

\begin{figure}
    \centering
    \includegraphics[width=0.67\linewidth]{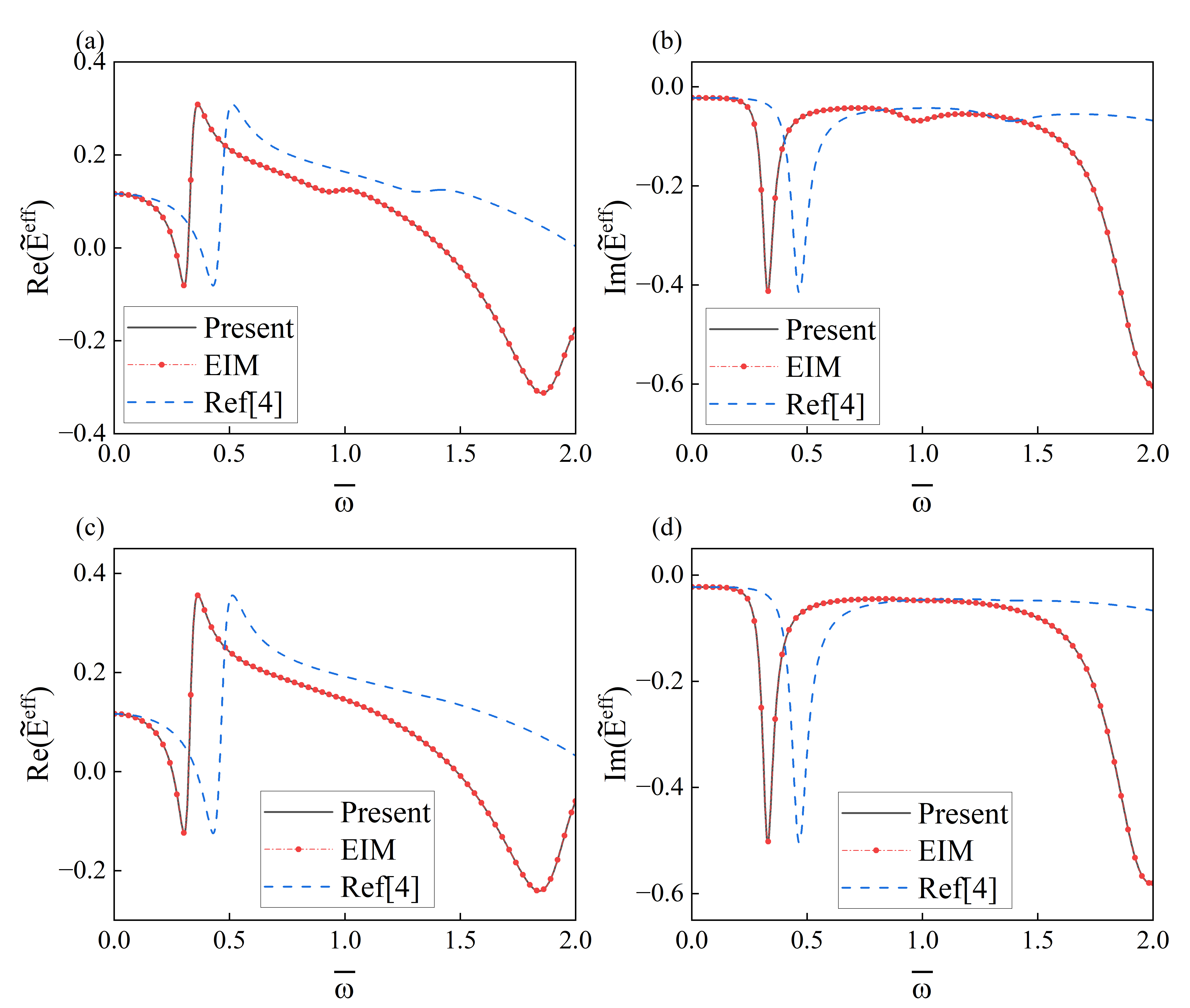}
    \caption{Comparison of the effective kernels obtained from \cite{Willis2012} (Ref [4]), the present re-derived exact homogenization method (Present), and the EIM homogenization method (EIM) with residual fields for the weight function $w_Y= 1$. 
    (a) $Re(\tilde{E}^\text{eff})$ and (b) $Im(\tilde{E}^\text{eff})$ with the admissible effective kernels; and (c) $Re(\tilde{E}^\text{eff})$ and (d) $Im(\tilde{E}^\text{eff})$ with the unique effective kernels. The normalized frequency $\overline{\omega} = L \omega / \sqrt{Re(E_2) / \rho_2}$, and the macroscopic wavenumber $\zeta = 2$.}
    \label{fig:unweight_2}
\end{figure}

\begin{figure}
    \centering
    \includegraphics[width=0.67\linewidth]{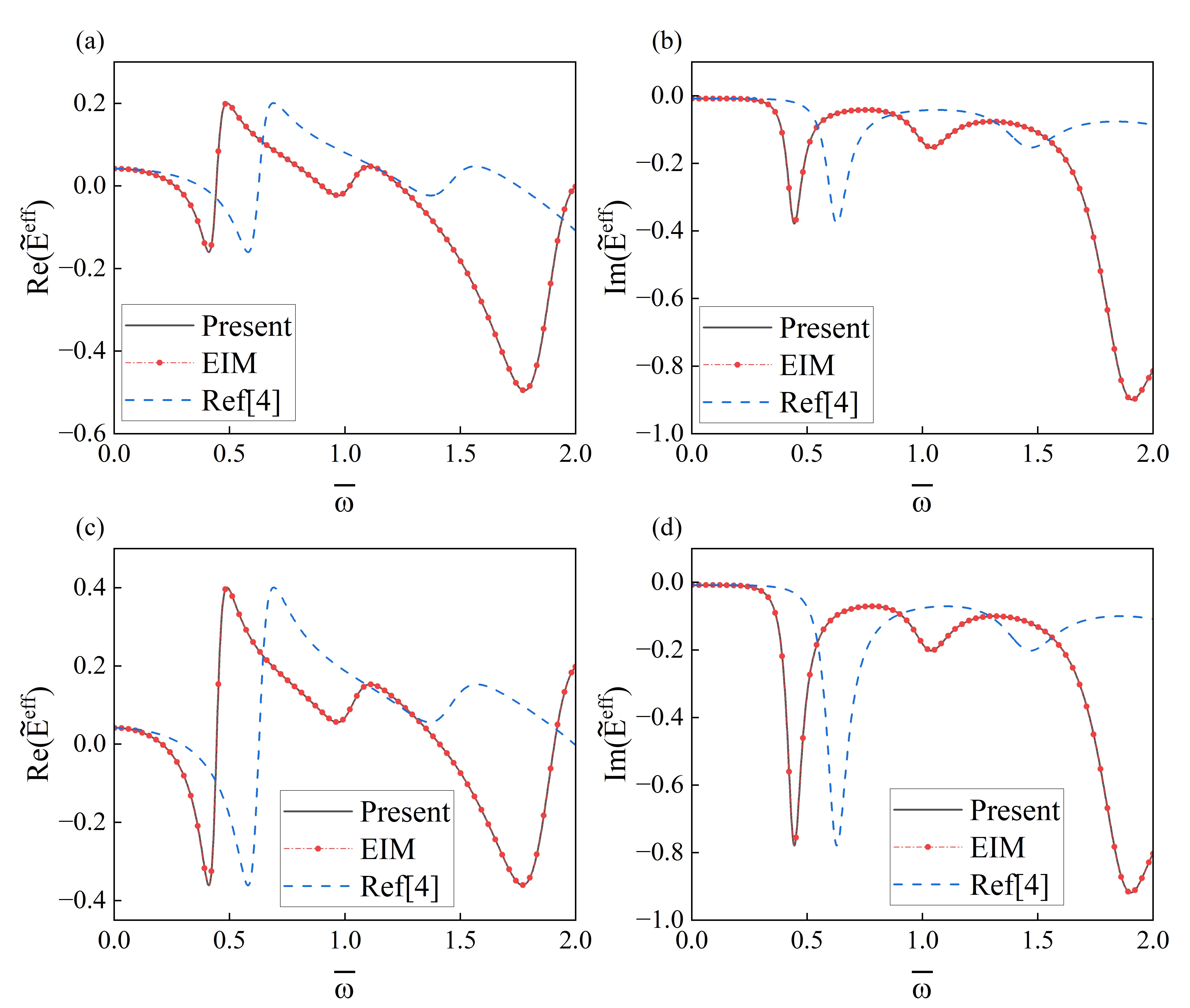}
    \caption{Comparison of the effective kernels obtained from \cite{Willis2012} (Ref [4]), the present re-derived exact homogenization method (Present), and the EIM homogenization method (EIM) with residual fields for the weight function $w_Y= 1/c_{2}$ in the second material phase. 
    (a) $Re(\tilde{E}^\text{eff})$ and (b) $Im(\tilde{E}^\text{eff})$ with the admissible effective kernels; and (c) $Re(\tilde{E}^\text{eff})$ and (d) $Im(\tilde{E}^\text{eff})$ with the unique effective kernels. The normalized frequency $\overline{\omega} = L \omega / \sqrt{Re(E_2) / \rho_2}$, and the macroscopic wavenumber $\zeta = 2$.}
    \label{fig:weight_2}
\end{figure}

Figs. \ref{fig:unweight_2} and \ref{fig:weight_2} provide the corresponding comparisons for Fig. 2 of \cite{Willis2012}. In particular, Figs. \ref{fig:unweight_2}(a) and (b) reproduce the real and imaginary parts shown in Fig. 2(a), whereas Figs. \ref{fig:unweight_2}(c) and (d) correspond to those in Fig. 2(c). Likewise, Figs. \ref{fig:weight_2}(a) and (b) correspond to Fig. 2(b), while Figs. \ref{fig:weight_2}(c) and (d) correspond to Fig. 2(d). 

\begin{figure}
    \centering
    \includegraphics[width=0.67\linewidth]{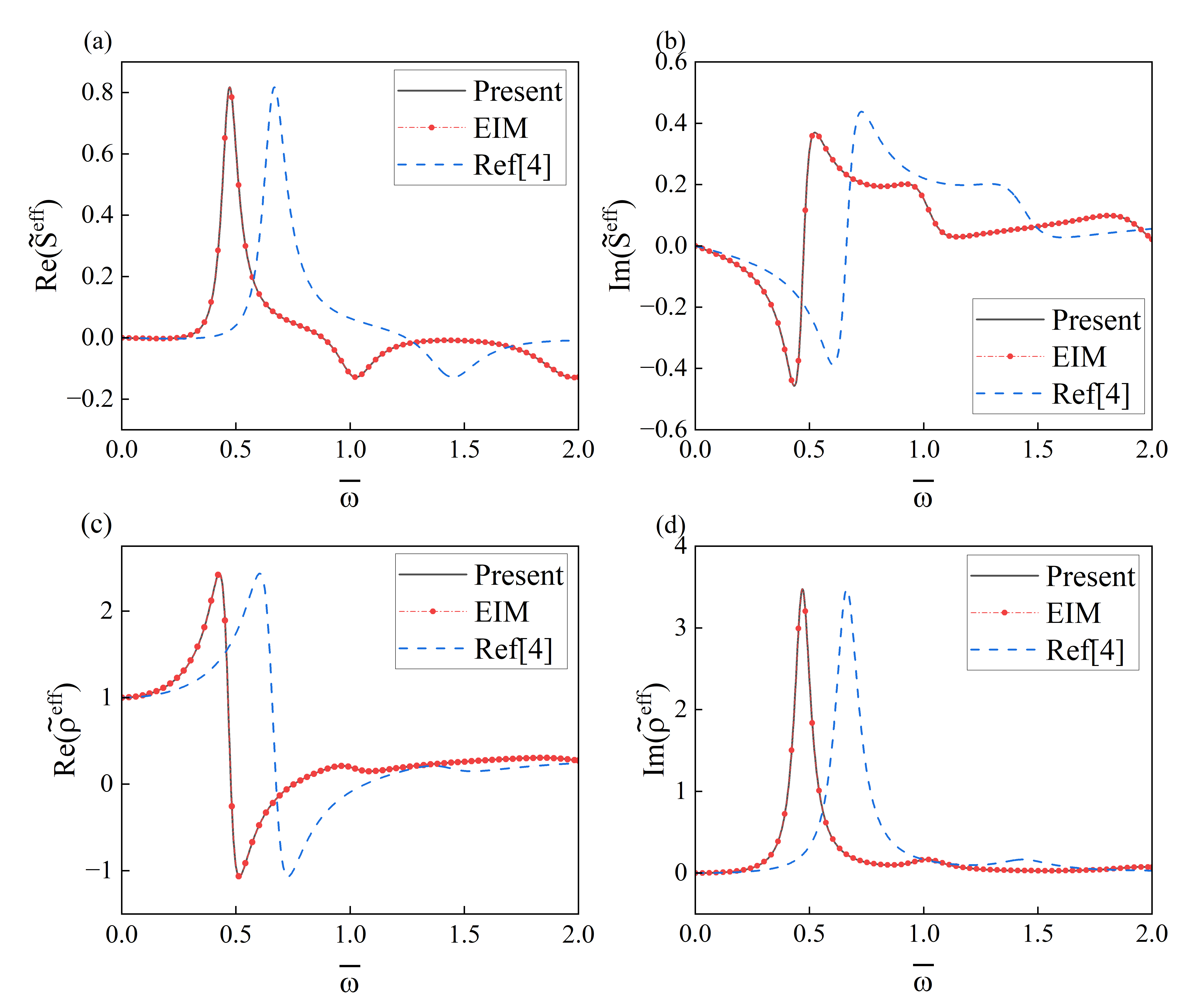}
    \caption{Comparison of the effective kernels obtained from \cite{Willis2012} (Ref [4]), the present re-derived exact homogenization method (Present), and the EIM homogenization method (EIM) with residual fields for the weight function $w_Y= 1/c_2$ in the second material phase. (a) $Re(\tilde{S}^\text{eff})$; (b) $Im(\tilde{S}^\text{eff})$; and (c) $Re(\tilde{\rho}^\text{eff})$ and (d) $Im(\tilde{\rho}^\text{eff})$. The normalized frequency $\overline{\omega} = L \omega / \sqrt{Re(E_2) / \rho_2}$, and the macroscopic wavenumber $\zeta = 0.1$.}
    \label{fig:2012_weight}
\end{figure}

Finally, Figs. \ref{fig:2012_weight}(a) and (b) reproduce the real and imaginary parts shown in Fig. 3(b) of \cite{Willis2012}, whereas Figs. \ref{fig:2012_weight}(c) and (d) correspond to those in Fig. 3(c).

\FloatBarrier

\end{document}